\documentclass[journal]{IEEEtran}

\usepackage{amsmath,epsfig,dsfont}
\usepackage{subfigure}
\usepackage{algorithm}
\usepackage{algorithmic}
\usepackage{amssymb,bm}
\usepackage{amsfonts}
\usepackage{stfloats}
\usepackage{multirow}
\ifCLASSINFOpdf
\else
\fi
\begin{document}

\title{Distributed Adaptive Networks: \\A Graphical Evolutionary Game-Theoretic View}
%
%
%
\author{Chunxiao~Jiang,~\IEEEmembership{Member,~IEEE,}
        Yan~Chen,~\IEEEmembership{Member,~IEEE,} and
        K.~J.~Ray~Liu,~\IEEEmembership{Fellow,~IEEE}
\thanks{Copyright (c) 2013 IEEE. Personal use of this material is permitted. However, permission to use this material for any other purposes must be obtained from the IEEE by sending a request to pubs-permissions@ieee.org.}
\thanks{Chunxiao Jiang is with Department of Electrical and Computer Engineering, University of Maryland, College Park, MD 20742, USA, and also
with Department of Electronic Engineering,
Tsinghua University, Beijing 100084, P. R. China (e-mail:
chx.jiang@gmail.com).}
\thanks{Yan Chen, and K. J. Ray Liu are with Department of Electrical and Computer Engineering,
University of Maryland, College Park, MD 20742, USA (e-mail: yan@umd.edu, kjrliu@umd.edu).}
}
\maketitle

\begin{abstract}
Distributed adaptive filtering has been considered as an effective approach for data processing and estimation over distributed networks. Most existing distributed adaptive filtering algorithms focus on designing different information diffusion rules, regardless of the nature evolutionary characteristic of a distributed network. In this paper, we study the adaptive network from the game theoretic perspective and formulate the distributed adaptive filtering problem as a graphical evolutionary game. With the proposed formulation, the nodes in the network are regarded as players and the local combiner of estimation information from different neighbors is regarded as different strategies selection. We show that this graphical evolutionary game framework is very general and can unify the existing adaptive network algorithms. Based on this framework, as examples, we further propose two error-aware adaptive filtering algorithms. Moreover, we use graphical evolutionary game theory to analyze the information diffusion process over the adaptive networks and evolutionarily stable strategy of the system. Finally, simulation results are shown to verify the effectiveness of our analysis and proposed methods.
\end{abstract}
%
\begin{IEEEkeywords}
Adaptive filtering, graphical evolutionary game, distributed estimation, adaptive networks, data diffusion.
\end{IEEEkeywords}
\section{Introduction}
Recently, the concept of adaptive filter network derived from the traditional adaptive filtering was emerging, where a group of nodes cooperatively estimate some parameters of interest from noisy measurements \cite{first}. Such a distributed estimation architecture can be applied to many scenarios, such as wireless sensor networks for environment monitoring, wireless Ad-hoc networks for military event localization, distributed cooperative sensing in cognitive radio networks and so on \cite{apply,cog}. Compared to the classical centralized architecture, the distributed one is not only more robust when the center node may be dysfunctional, but also more flexible when the nodes are with mobility. Therefore, distributed adaptive filter network has been considered as an effective approach for the implementation of data fusion, diffusion and processing over distributed networks \cite{book}.

In a distributed adaptive filter network, at every time instant $t$, node $i$ receives a set of data $\{d_i(t), \bm u_{i,t}\}$ that satisfies a linear regression model as follow
\begin{equation}
d_i(t)=\bm u_{i,t}\bm w^0+v_i(t),
\end{equation}
where $\bm w^0$ is a deterministic but unknown $M\times 1$ vector, $d_i(t)$ is a scalar measurement of some random process $\bm d_i$, $\bm u_{i,t}$ is the $1\times M$ regression vector at time $t$ with zero mean and covariance matrix $\bm R_{u_i}=\mathbb{E}\big(u^{*}_{i,t}u_{i,t}\big)>0$, and $v_i(t)$ is the random noise signal at time $t$ with zero mean and variance $\sigma^2_i$. Note that the regression data $\bm u_{i,t}$ and measurement process $\bm d_i$ are temporally white and spatially independent, respectively and mutually. The objective for each node is to use the data set $\{d_i(t), \bm u_{i,t}\}$ to estimate parameter $\bm w^0$.

In the literatures, many distributed adaptive filtering algorithms have been proposed for the estimation of parameter $\bm w^0$. The incremental algorithms, in which node $i$ updates $\bm w$, i.e., the estimation of $\bm w^0$, through combining the observed data sets of itself and node $i-1$, were proposed, e.g., the incremental LMS algorithm \cite{incre}. Unlike the incremental algorithms, the diffusion algorithms allow node $i$ to combine the data sets from all neighbors, e.g., diffusion LMS \cite{lms,lms2} and diffusion RLS \cite{rls}. Besides, the projection-based adaptive filtering algorithms were summarized in \cite{spm}, e.g., the projected subgradient algorithm \cite{projection1} and the combine-project-adapt algorithm \cite{projection2}. In \cite{mobile}, the authors considered the node's mobility and analyzed the mobile adaptive networks.

While achieving promising performance, these traditional distributed adaptive filtering algorithms mainly focused on designing different information combination rules or diffusion rules among the neighborhood by utilizing the network topology information and/or nodes' statistical information. For example, the relative degree rule considers the degree information of each node \cite{rls}, and the relative degree-variance rule further incorporates the variance information of each node \cite{lms}. However, most of the existing algorithms are somehow intuitively designed to achieve some specific objective, sort of like bottom-up approaches to the distributed adaptive networks. There is no existing work that offers a design philosophy to explain why combination and/or diffusion rules are developed and how they are related in a unified view. Is there a general framework that can reveal the relationship among the existing rules and provide fundamental guidance for better design of distributed adaptive filtering algorithms? In our quest to answer the question, we found that in essence the parameter updating process in distributed adaptive networks follows similarly the evolution process in natural ecological systems. Therefore, based on the graphical evolutionary game, in this paper, we propose a general framework that can offer a unified view of existing distributed adaptive algorithms, and provide possible clues for new future designs. Unlike the traditional bottom-up approaches that focus on some specific rules, our framework provide a top-down design philosophy to understand the fundamental relationship of distributed adaptive networks.

The main contributions of this paper are summarized as follows.
\begin{enumerate}
\item We propose a graphical evolutionary game theoretic framework for the distributed adaptive networks, where nodes in the network are regarded as players and the local combination of estimation information from different neighbors is regarded as different strategies selection. We show that the proposed graphical evolutionary theoretic framework can unify existing adaptive filtering algorithms as special cases.
\item Based on the proposed framework, as examples, we further design two simple error-aware distributed adaptive filtering algorithms. When the noise variance is unknown, our proposed algorithm can achieve similar performance compared with existing algorithms but with lower complexity, which immediately shows the advantage of the proposed general framework.
\item Using the graphical evolutionary game theory, we analyze the information diffusion process over the adaptive network, and derive the diffusion probability of information from good nodes.
\item We prove that the strategy of using information from good nodes is evolutionarily stable strategy either in complete graphs or incomplete graphs.
\end{enumerate}

The rest of this paper is organized as follows. We summarize the existing works in Section II. In Section III, we describe in details how to formulate the distributed adaptive filtering problem as a graphical evolutionary game. We then discuss the information diffusion process over the adaptive network in Section IV, and further analyze the evolutionarily stable strategy in Section V. Simulation results are shown in Section VI. Finally, we draw conclusions in Section VII.

\section{Related Works}

\begin{figure}[!t]
  \centerline{\epsfig{figure=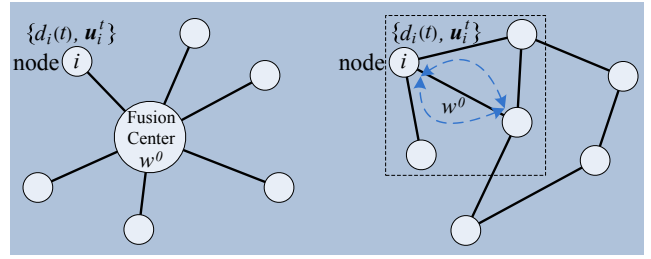,width=8.5cm}}
  \caption{Left: centralized model. Right: distributed model.}\label{fig1}
\end{figure}

Let us consider an adaptive filter network with $N$ nodes. If there is a fusion center that can collect information from all nodes, then global (centralized) optimization methods can be used to derive the optimal updating rule for the parameter $\bm w$, where $\bm w$ is a deterministic but unknown $M \times 1$ vector for estimation, as shown in the left part of Fig.\,\ref{fig1}. For example, in the global LMS algorithm, the parameter updating rule can be written as \cite{lms}
\begin{equation}
\bm w_{t+1}=\bm w_t+\mu\sum\limits_{i=1}^N \bm u^{*}_{i,t}\Big(d_i(t)-\bm u_{i,t}\bm w_t\Big),\label{global}
\end{equation}
where $\mu$ is the step size and $\{\cdot\}^*$ denotes complex conjugation operation. With (\ref{global}), we can see that the centralized LMS algorithm requires the information of $\{d_i(t), \bm u_{i,t}\}$ across the whole network, which is generally impractical. Moreover, such a centralized architecture highly relies on the fusion center and will collapse when the fusion center is dysfunctional or some data links are disconnected.

If there is no fusion center in the network, then each node needs to exchange information with the neighbors to update the parameter as shown in the right part of Fig.\,\ref{fig1}. In the literature, several distributed adaptive filtering algorithms have been introduced, such as distributed incremental algorithms \cite{incre}, distributed LMS \cite{lms}, \cite{lms2}, and projection-based algorithms \cite{projection1}, \cite{projection2}. These distributed algorithms are based on the classical adaptive filtering algorithms, where the difference is that nodes can use information from neighbors to estimate the parameter $\bm w^0$. Taking one of the distributed LMS algorithms, Adapt-then-Combine Diffusion LMS (ATC) \cite{lms}, as an example, the parameter updating rule for node $i$ is
\begin{eqnarray}
\left\{ \begin{array}{l}
\bm \chi_{i,t+1}=\bm w_{i,t}+\mu_i\sum\limits_{j\in\mathcal N_i}o_{ij} \bm u^{*}_{j,t}\Big(d_j(t)-\bm u_{j,t}\bm w_{j,t}\Big), \vspace{2mm}\\
\bm w_{i,t+1}=\sum\limits_{j\in\mathcal N_i}a_{ij}\bm\chi_{j,t+1},
\end{array}\right.
\end{eqnarray}
where $\mathcal N_i$ denotes the neighboring nodes set of node $i$ (including node $i$ itself), $o_{ij}$ and $a_{ij}$ are linear weights satisfying the following conditions
\begin{eqnarray}
\left\{ \begin{array}{l}
o_{ij}=a_{ij}=0, \quad \mbox{if}\ j\notin \mathcal N_i, \vspace{2mm}\\
\sum\limits_{j=1}^N o_{ij}=1, \quad \sum\limits_{j=1}^N a_{ij}=1.
\end{array}\right.
\end{eqnarray}

In a practical scenario, since the exchange of full raw data $\{d_i(t), \bm u_{i,t}\}$ among neighbors is costly, the weight $o_{ij}$ is usually set as $o_{ij}=0$, if $j\neq i$, as in \cite{lms}. In such a case, for node $i$ with degree $n_i$ (including node $i$ itself, i.e., the cardinality of set $\mathcal N_i$) and neighbour set $\{i_1,i_2,\dots,i_{n_i}\}$, we can write the general parameter updating rule as
\begin{eqnarray}
\bm{w}_{i,t+1}\!\!\!&=&\!\!\!A_{i,t+1}\Big(F(\bm{w}_{i_1,t}),F(\bm{w}_{i_2,t}),...,F(\bm{w}_{i_{n_i},t})\Big),\nonumber\\
\!\!\!&=&\!\!\!\sum_{j\in \mathcal N_i}A_{i,t+1}(j)F(\bm{w}_{j,t}),\label{general}
\end{eqnarray}
where $F(\cdot)$ can be any adaptive filtering algorithm, e.g. $F(\bm{w}_{i,t})=\bm{w}_{i,t}+\mu \bm{u}_{i,t}^{*}(d_i(t)-\bm{u}_{i,t}\bm{w}_{i,t})$ for the LMS algorithm, $A_{i,t+1}(\cdot)$ represents some specific linear combination rule. The (\ref{general}) gives a general form of existing distributed adaptive filtering algorithms, where the combination rule $A_{i,t+1}(\cdot)$ mainly determines the performance. Table~\ref{table1} summarizes the existing combination rules, where for all rules $A_{i,t+1}(j)=0$, if $j\notin \mathcal N_i$.

\begin{table}[!t]\renewcommand{\arraystretch}{2}
    \caption{Different Combination Rules.}\label{table1}\vspace{2mm}
    \begin{center}
    \begin{tabular}{|c|c|}
    \hline
    \textbf{Name} & \textbf{Rule}: $A_i(j)=$\\ \hline
    Uniform \cite{projection2}\cite{uniform}& $
\frac{1}{n_i}, \ \mbox{for all}\ j\in\mathcal N_i
$ \\ \hline
    Maximum degree \cite{rls}\cite{maximum} & $
\left\{\begin{array}{ll}
\!\!\!\!\frac{1}{N},&\!\!\!\!\mbox{for} \ j\neq i, \vspace{0mm}\\
\!\!\!\!1-\frac{n_i-1}{N},&\!\!\!\!\mbox{for} \  j= i .
\end{array}\right.$ \\ \hline
    Laplacian \cite{laplacian}\cite{laplacian1}& $
\left\{\begin{array}{ll}
\!\!\!\!\frac{1}{n_{\mbox{\tiny{max}}}},&\!\!\!\!\mbox{for} \ j\neq i \vspace{0mm}\\
\!\!\!\!1-\frac{n_i-1}{n_{\mbox{\tiny{max}}}},&\!\!\!\!\mbox{for} \  j= i .
\end{array}\right.$ \\ \hline
    Relative degree \cite{rls}& $\frac{n_j}{\sum_{k\in\mathcal N_i} n_k},\ \mbox{for all}\ j\in\mathcal N_i$ \\ \hline
    Relative degree-variance \cite{lms}& $\frac{n_j\sigma^{-2}_{j}}{\sum_{k\in\mathcal N_i}n_k\sigma^{-2}_{k}},\ \mbox{for all}\ j\in\mathcal N_i$ \\ \hline
    Metropolis \cite{laplacian1}\cite{two}& $
\left\{\begin{array}{ll}
\!\!\!\!\frac{1}{\max\{|\mathcal N_i|,|\mathcal N_j|\}},&\!\!\!\!\mbox{for} \ j\neq i, \vspace{0mm}\\
\!\!\!\!1-\sum_{k\neq i}A_i(k),&\!\!\!\!\mbox{for} \  j= i .
\end{array}\right.$ \\ \hline
Hastings \cite{two}& $
\left\{\begin{array}{ll}
\!\!\!\!\frac{\sigma^{2}_{j}}{\max\{|\mathcal N_i|\sigma^{2}_{i},|\mathcal N_j|\sigma^{2}_{j}\}},&\!\!\!\!\mbox{for} \ j\neq i, \vspace{0mm}\\
\!\!\!\!1-\sum_{k\neq i}A_i(k),&\!\!\!\!\mbox{for} \  j= i .
\end{array}\right.$ \\ \hline

    \end{tabular}
    \end{center}
\end{table}

From Table~\ref{table1}, we can see that the weights of the first four combination rules are purely based on the network topology. The disadvantage of such topology-based rules is that, they are sensitive to the spatial variation of signal and noise statistics across the network. The relative degree-variance rule shows better mean-square performance than others, which, however, requires the knowledge of all neighbors' noise variances. As discussed in Section I, all these distributed algorithms are only focusing on designing the combination rules. Nevertheless, a distributed network is just like a natural ecological system and the nodes are just like individuals in the system, which may spontaneously follow some nature evolutionary rules, instead of some specific artificially predefined rules. Besides, although various kinds of combination rules have been developed, there is no general framework which can reveal the unifying fundamentals of distributed adaptive filtering problems. In the sequel, we will use graphical evolutionary game theory to establish a general framework to unify existing algorithms and give insights of the distributed adaptive filtering problem.

\section{Graphical Evolutionary Game Formulation}

\subsection{Introduction of Graphical Evolutionary Game}


Evolutionary game theory (EGT) is originated from the study of ecological biology \cite{evol}, which differs from the classical game theory by emphasizing more on the dynamics and stability of the whole population's strategies \cite{evol2}, instead of only the property of the equilibrium. EGT has been widely used to model users' behaviors in image processing \cite{yan}, as well as communication and networking area \cite{bookliu}\cite{reviewwang}, such as congestion control \cite{cong}, cooperative sensing \cite{cs}, cooperative peer-to-peer (P2P) streaming \cite{p2p} and dynamic spectrum access \cite{joint}. In these literatures, evolutionary game has been shown to be an effective approach to model the dynamic social interactions among users in a network.

EGT is an effective approach to study how a group of players converges to a stable equilibrium after a period of strategic interactions. Such an equilibrium strategy is defined as the Evolutionarily Stable Strategy (ESS). For an evolutionary game with $N$ players, a strategy profile $\mathbf{a}^*=(a_1^*,...,a_N^*)$, where $a^*_i\in\mathcal X$ and $\mathcal X$ is the action space, is an ESS if and only if, $\forall \mathbf {a} \neq \mathbf a^*$, $\mathbf a^*$ satisfies following \cite{evol2}:
\begin{eqnarray}
\!\!\!\!\!\!\!\!\!\!\!\!\!\!\!\!\!\!&&1)\ U_i(a_i,\ \mathbf a^*_{-i})\le U_i(a_i^*,\ \mathbf a^*_{-i}),\quad\quad\quad\quad\quad\quad\quad\\
\!\!\!\!\!\!\!\!\!\!\!\!\!\!\!\!\!\!&&2)\ \mbox{if} \ U_i(a_i,\ \mathbf a^*_{-i})=U_i(a_i^*,\ \mathbf a^*_{-i}),\nonumber\\
 &&\quad\quad\ U_i(a_i,\ \mathbf a_{-i})<U_i(a_i^*,\ \mathbf a_{-i}),\label{ess}
\end{eqnarray}
where $U_i$ stands for the utility of player $i$ and $\mathbf a_{-i}$ denotes the strategies of all players other than player $i$. We can see that the first condition is the Nash equilibrium (NE) condition, and the second condition guarantees the stability of the strategy. Moreover, we can also see that a strict NE is always an ESS. If all players adopt the ESS, then no mutant strategy could invade the population under the influence of natural selection. Even if a small part of players may not be rational and take out-of-equilibrium strategies, ESS is still a locally stable state.

Let us consider an evolutionary game with $m$ strategies $\mathcal X=\{1,2,...,m\}$. The utility matrix, $U$, is an $m\times m$ matrix, whose entries, $u_{ij}$, denote the utility for strategy $i$ versus strategy $j$. The population fraction of strategy $i$ is given by $p_i$, where $\sum_{i=1}^m p_i=1$. The fitness of strategy $i$ is given by $f_i=\sum_{j=1}^m p_ju_{ij}$. For the average fitness of the whole population, we have $\phi=\sum_{i=1}^mp_if_i$. The Wright-Fisher model has been widely adopted to let a group of players converge to the ESS \cite{fisher}, where the strategy updating equation for each player can be written as
\begin{equation}
p_i(t+1)=\frac{p_i(t)f_i(t)}{\phi(t)}.\label{fish}
\end{equation}
Note that one assumption in the Wright-Fisher model is that when the total population is sufficiently large, the fraction of players using strategy $i$ is equal to the probability of one individual player using strategy $i$. From (\ref{fish}), it can be seen that the strategy updating process in the evolutionary game is similar to the parameter updating process in adaptive filter problem. It is intuitive that we can use evolutionary game to formulate the distributed adaptive filter problem.

\begin{figure}[!t]
  \centerline{\epsfig{figure=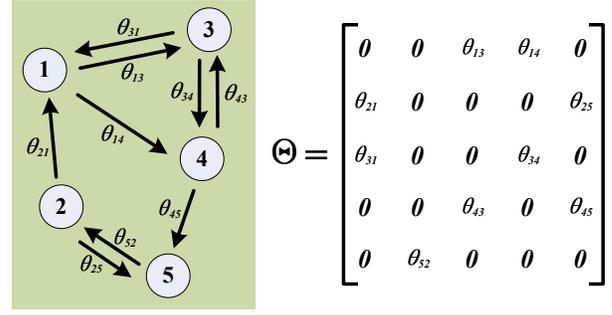,width=8cm}}
  \caption{Graphical evolutionary game model.}\label{fig2}
\end{figure}

The classical evolutionary game theory considers a population of $M$ individuals in a complete graph. However, in many scenarios, players' spatial locations may lead to an incomplete graph structure. Graphical evolutionary game theory is introduced to study the strategies evolution in such a finite structured population \cite{geg}, where each vertex represents a player and each edge represents the reproductive relationship between valid neighbors, i.e., $\theta_{ij}$ denotes the probability that the strategy of node $i$ will replace that of node $j$, as shown in Fig.\,\ref{fig2}. Graphical EGT focuses on analyzing the ability of a mutant gene to overtake a group of finite structured residents. One of the most important research issues in graphical EGT is how to compute the fixation probability, i.e., the probability that the mutant will eventually overtake the whole structured population \cite{gegreview}. In the following, we will use graphical EGT to formulate the dynamic parameter updating process in a distributed adaptive filter network.

\subsection{Graphical Evolutionary Game Formulation}

In graphical EGT, each player updates strategy according to his/her fitness after interacting with neighbors in each round. Similarly, in distributed adaptive filtering, each node updates its parameter $\bm{w}$ through incorporating the neighbors' information. In such a case, we can treat the nodes in a distributed filter network as players in a graphical evolutionary game. For node $i$ with $n_i$ neighbors, it has $n_i$ pure strategies $\{i_1,i_2,...,i_{n_i}\}$, where strategy $j$ means updating $\bm w_{i,t+1}$ using the updated information from its neighbor $j$, $A_{i,t+1}(j)$. We can see that (\ref{general}) represents the adoption of mixed strategy. In such a case, the parameter updating in distributed adaptive filter network can be regarded as the strategy updating in graphical EGT. Table~\ref{table3} summarizes the correspondence between the terminologies in graphical EGT and those in distributed adaptive network.

\begin{table}[!t]\renewcommand{\arraystretch}{2}
    \caption{Correspondence Between Graphical EGT and Distributed Adaptive Network.}\label{table3}\vspace{2mm}
    \begin{center}
    \begin{tabular}{|c|c|}
    \hline
    \textbf{Graphical EGT} & \textbf{Distributed adaptive network}\\ \hline
    $N$ Players &$N$ Nodes in the network\\ \hline
    \multirow{2}{3.6cm}{Pure strategy of player $i$ with $n_i$ neighbors $\{i_1,i_2,...,i_{n_i}\}$}& \multirow{2}{4.2cm}{Node $i$ combines information from one of its neighbors $\{i_1,i_2,...,i_{n_i}\}$} \\ &\\ \hline
    \multirow{2}{3.6cm}{Mixed strategy of player $i$ with $n_i$ neighbors $\{p_1,p_2,...,p_{n_i}\}$} & \multirow{2}{3.5cm}{Node $i$'s combiner (Weight) $\{A_{i}(1),A_{i}(2),...,A_{i}(n_i)\}$} \\ &\\ \hline
    Mixed strategy update of player $i$ & Combiner update of node $i$\\ \hline
    Equilibrium& Convergence network state\\ \hline
    \end{tabular}
    \end{center}
\end{table}

\begin{figure}[!t]
\begin{minipage}[t]{1\linewidth}
  \centering
  \centerline{\epsfig{figure=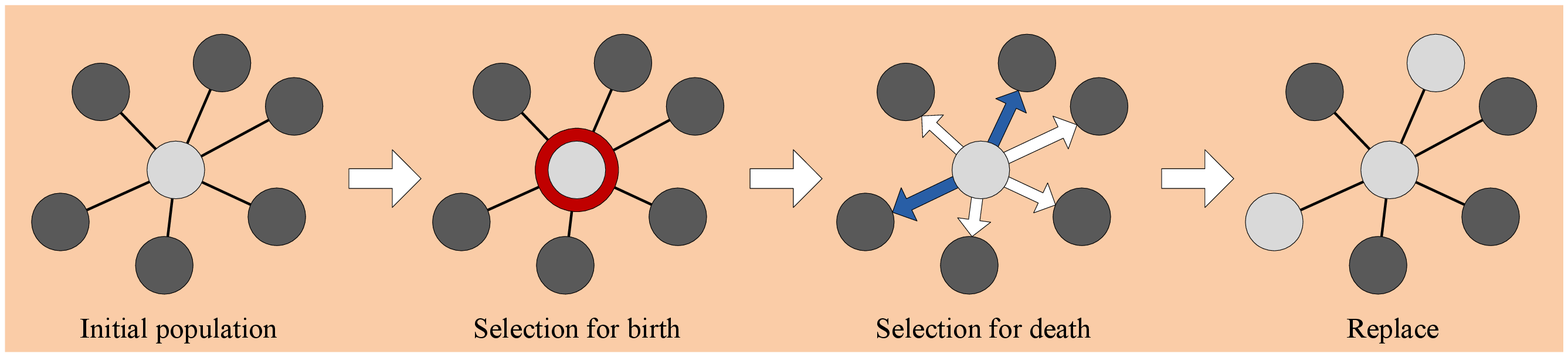,width=8.5cm}}
  \centerline{\scriptsize{(a) BD update rule.}}\vspace{0.3cm}
\end{minipage}
\begin{minipage}[t]{1\linewidth}
  \centering
  \centerline{\epsfig{figure=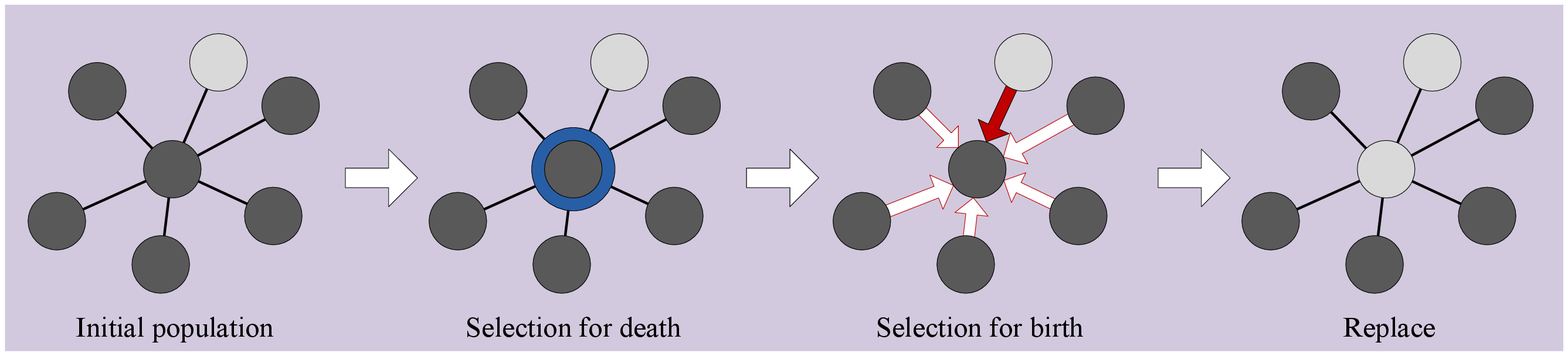,width=8.5cm}}
  \centerline{\scriptsize{(b) DB update rule.}}\vspace{0.3cm}
\end{minipage}
\begin{minipage}[t]{1\linewidth}
  \centering
  \centerline{\epsfig{figure=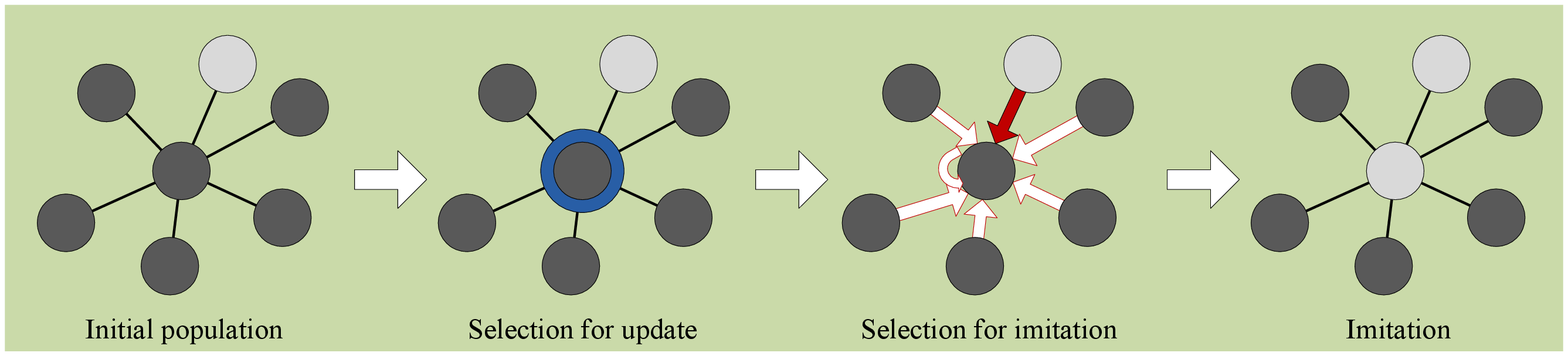,width=8.5cm}}
  \centerline{\scriptsize{(c) IM update rule.}}\vspace{0.2cm}
\end{minipage}
\caption{Three different update rules, where death selections are shown in dark blue and birth selections are shown in red.}\label{fig3}
\end{figure}

We first discuss how players' strategies are updated in graphical EGT, which is then applied to the parameter updating in distributed adaptive filtering.
In graphical EGT, the fitness of a player is locally determined from interactions with all adjacent players, which is defined as \cite{fitness}
\begin{equation}
f=(1-\alpha)\cdot B+\alpha\cdot U,\label{fitness}
\end{equation}
where $B$ is the baseline fitness, which represents the player's inherent property. For example, in a distributed adaptive network, a node's baseline fitness can be interpreted as the quality of its noise variance. $U$ is the player's utility which is determined by the predefined utility matrix. The parameter $\alpha$ represents the selection intensity, i.e., the relative contribution of the game to fitness. The case $\alpha\rightarrow 0$ represents the limit of weak selection \cite{weak}, while $\alpha=1$ denotes strong selection, where fitness equals utility. There are three different strategy updating rules for the evolution dynamics, called as birth-death (BD), death-birth (DB) and imitation (IM) \cite{rd}.
\begin{itemize}
\item
BD update rule: a player is chosen for reproduction with the probability being proportional to fitness (Birth process). Then, the chosen player's strategy replaces one neighbor's strategy uniformly (Death process), as shown in Fig.\,\ref{fig3}-(a).
\item
DB update rule: a random player is chosen to abandon his/her current strategy (Death process). Then, the chosen player adopts one of his/her neighbors' strategies with the probability being proportional to their fitness (Birth process), as shown in Fig.\,\ref{fig3}-(b).
\item
IM update rule: each player either adopts the strategy of one neighbor or remains with his/her current strategy, with the probability being proportional to fitness, as shown in Fig.\,\ref{fig3}-(c).
\end{itemize}

These three kinds of strategy updating rules can be matched to three different kinds of parameter updating algorithms in distributed adaptive filtering. Suppose that there are $N$ nodes in a structured network, where the degree of node $i$ is $n_i$. We use $\mathcal N$ to denote the set of all nodes and $\mathcal N_i$ to denote the neighborhood set of node $i$, including node $i$ itself.

For the BD update rule, the probability that node $i$ adopts strategy $j$, i.e., using updated information from its neighbor node $j$, is
\begin{equation}
P_j=\frac{f_j}{\sum_{k\in\mathcal{N}} f_k}\frac{1}{n_j},\label{pjbd}
\end{equation}
where the first term $\frac{f_j}{\sum_{k\in\mathcal{N}} f_k}$ is the probability that the neighboring node $j$ is chosen to reproduction, which is proportional to its fitness $f_j$, and the second term $\frac{1}{n_j}$ is the probability that node $i$ is chosen for adopting strategy $j$. Note that the network topology information ($n_j$) is required to calculate (\ref{pjbd}).
In such a case, the equivalent parameter updating rule for node $i$ can be written by
\begin{eqnarray}
\bm w_{i,t+1}\!\!\!&=&\!\!\!\sum\limits_{j\in\mathcal{N}_i\backslash \{i\}}\bigg(\frac{f_j}{\sum_{k\in\mathcal{N}} f_k}\frac{1}{n_j}\bigg)F(\bm w_{j,t})+\nonumber\\
\!\!\!&&\!\!\!\Bigg(1-\sum\limits_{j\in\mathcal{N}_i\backslash \{i\}}\bigg(\frac{f_j}{\sum_{k\in\mathcal{N}} f_k}\frac{1}{n_j}\bigg)\Bigg)F(\bm w_{i,t}).\label{bd}
\end{eqnarray}
Similarly, for the DB updating rule, we can obtain the corresponding parameter updating rule for node $i$ as
\begin{eqnarray}
\!\!\!\bm w_{i,t+1}\!\!\!&=&\!\!\!\frac{1}{n_i}\sum_{j\in\mathcal{N}_i\backslash \{i\}}\bigg(\frac{f_j}{\sum_{k\in\mathcal{N}_i} f_k}\bigg)F(\bm w_{j,t})+\nonumber\\
\!\!\!&&\!\!\!\Bigg(1-\frac{1}{n_i}\sum_{j\in\mathcal{N}_i\backslash \{i\}}\bigg(\frac{f_j}{\sum_{k\in\mathcal{N}_i} f_k}\bigg)\Bigg)F(\bm w_{i,t}).\label{db}
\end{eqnarray}
For the IM updating rule, we have
\begin{equation}
\bm w_{i,t+1}=\sum\limits_{j\in\mathcal{N}_i}\bigg(\frac{f_j}{\sum_{k\in\mathcal{N}_i} f_k}\bigg)F(\bm w_{j,t}).\label{im}
\end{equation}
Note that (\ref{bd}), (\ref{db}) and (\ref{im}) are expected outcome of BD, DB and IM updated rules, which can be referred in [35], [37].

The performance of adaptive filtering algorithm is usually evaluated by two measures: mean-square deviation (MSD) and excess-mean-square error (EMSE), which are defined as
\begin{eqnarray}
\mbox{MSD}\!\!\!&=&\!\!\!\mbox{E}||\bm w_t-\bm w^0||^2,\\
\mbox{EMSE}\!\!\!&=&\!\!\!\mbox{E}\left|\bm u_t(\bm w_{t-1}-\bm w^0)\right|^2.
\end{eqnarray}
Using (\ref{bd}), (\ref{db}) and (\ref{im}), we can calculate the network MSD and EMSE of these three update rules according to \cite{lms}.

\subsection{Relationship to Existing Distributed Adaptive Filtering Algorithms}

In Section II, we have summarized the existing distributed adaptive filtering algorithms in (\ref{general}) and Table~\ref{table1}. In this subsection, we will show that all these algorithms are the special cases of the IM update rule in our proposed graphical EGT framework. Compare (\ref{general}) and (\ref{im}), we can see that different fitness definitions are corresponding to different distributed adaptive filtering algorithms in Table~\ref{table1}. For the uniform rule, the fitness can be uniformly defined as $f_i=1$ and using the IM update rule, we have
\begin{equation}
\bm w_{i,t+1}=\sum\limits_{j\in\mathcal{N}_i}\frac{1}{n_i}F(\bm w_{j,t}),
\end{equation}
which is equivalent to the uniform rule in Table~\ref{table1}. Here, the definition of $f_i=1$ means the adoption of fixed fitness and weak selection ($\alpha<<1$).
For the Laplacian rule, when updating the parameter of node $i$, the fitness of nodes in $\mathcal N_i$ can be defined as
\begin{eqnarray}
f_j=\left\{\begin{array}{ll}
1,&\mbox{for} \ j\neq i, \vspace{0mm}\\
n_{\mbox{\scriptsize max}}-n_i+1,&\mbox{for} \  j= i .
\end{array}\right.\label{laplacian}
\end{eqnarray}
From (\ref{laplacian}), we can see that each node gives more weight to the information from itself through enhancing its own fitness.
Similarly, for the Relative-degree-variance rule, the fitness can be defined as
\begin{equation}
f_j=n_j\sigma_j^{-2},\quad \mbox{for all} \ j\in\mathcal N_i.
\end{equation}

\begin{table}[!t]\renewcommand{\arraystretch}{1.5}
    \caption{Different Fitness Definitions.}\label{table2}\vspace{2mm}
    \begin{center}
    \begin{tabular}{|c|c|}
    \hline
    \textbf{Name} & \textbf{Fitness}: $f_j=$\\ \hline
    Uniform \cite{projection2}\cite{uniform}& $
1, \ \mbox{for all}\ j\in\mathcal N_i
$ \\ \hline
    Maximum& \multirow{2}{3cm}{$
\left\{\begin{array}{ll}
\!\!1,&\!\!\!\!\mbox{for} \ j\neq i,\\
\!\!N-n_i+1,&\!\!\!\!\mbox{for} \  j= i .
\end{array}\right.$}\\
degree \cite{rls}\cite{maximum}& \\\hline
    Laplacian \cite{laplacian}\cite{laplacian1}& $
\left\{\begin{array}{ll}
\!\!1,&\!\!\!\!\mbox{for} \ j\neq i \\
\!\!n_{\mbox{\scriptsize{max}}}-n_i+1,&\!\!\!\!\mbox{for} \  j= i .
\end{array}\right.$ \\ \hline
    Relative degree \cite{rls}& $n_j,\ \mbox{for all}\ j\in\mathcal N_i$ \\ \hline
    Relative & \multirow{2}{2.8cm}{$n_j\sigma^{-2}_{j},\ \mbox{for all}\ j\in\mathcal N_i$} \\degree-variance \cite{lms}& \\ \hline
    Metropolis \cite{laplacian1}\cite{two}& $\begin{pmatrix} 1-\sum_{k\neq i}A_i(k)&\frac{1}{\max\{|\mathcal N_i|,|\mathcal N_j|\}} \\ \frac{1}{\max\{|\mathcal N_i|,|\mathcal N_j|\}}&1-\sum_{k\neq j}A_j(k) \end{pmatrix}$\\ \hline
    Hastings \cite{two}& $\begin{pmatrix} 1-\sum_{k\neq i}A_i(k) \!\!\!&\!\!\! \frac{\sigma^{2}_{(i,j)}}{\max\{|\mathcal N_i|\sigma^{2}_{i},|\mathcal N_j|\sigma^{2}_{j}\}}\\
    \frac{\sigma^{2}_{(j,i)}}{\max\{|\mathcal N_i|\sigma^{2}_{i},|\mathcal N_j|\sigma^{2}_{j}\}}\!\!\!&\!\!\!1-\sum_{k\neq j}A_j(k) \end{pmatrix}$\\\hline
    \end{tabular}
    \end{center}
\end{table}

For the metropolis rule and Hastings rule, the corresponding fitness definitions are based on strong selection model ($\alpha\rightarrow 1$), where utility plays a dominant role in (\ref{fitness}). For the metropolis rule, the utility matrix of nodes can be defined as
\begin{eqnarray}
    \begin{tabular}{ccccc}
    && Node $i$ &Node $j\neq i$&\vspace*{-8pt}\\
    &&&&\\
    \!\!\!\!Node $i$ \!\!\!\!\!\!\!\!& \multirow{3}{0.0001cm}{\Bigg(}& $1-\sum_{k\neq i}A_i(k)$&$\frac{1}{\max\{|\mathcal N_i|,|\mathcal N_j|\}}$&\!\!\!\!\!\!\!\!\!\!\multirow{3}{0.0001mm}{\Bigg)\!\!\!\!}\vspace*{-4pt}\\
    &&&&\\
    \!\!\!\!Node $j\neq i$ \!\!\!\!\!\!\!\!& & $\frac{1}{\max\{|\mathcal N_i|,|\mathcal N_j|\}}$&$1-\sum_{k\neq j}A_j(k)$&
    \end{tabular}\label{metropolis}
\end{eqnarray}
For the Hastings rule, the utility matrix can be defined as
\begin{eqnarray}
    \begin{tabular}{ccccc}
    && Node $i$ &Node $j\neq i$&\vspace*{-8pt}\\
    &&&&\\
    \!\!\!\!Node $i$ \!\!\!\!\!\!\!\!& \multirow{3}{0.0001cm}{\Bigg(}& $1-\sum_{k\neq i}A_i(k)$&$\frac{\sigma^{2}_{(i,j)}}{\max\{|\mathcal N_i|\sigma^{2}_{i},|\mathcal N_j|\sigma^{2}_{j}\}}$&\!\!\!\!\!\!\!\!\!\!\multirow{3}{0.0001mm}{\Bigg)\!\!\!\!}\vspace*{-8pt}\\
    &&&&\\
    \!\!\!\!Node $j\neq i$ \!\!\!\!\!\!\!\!& & $\frac{\sigma^{2}_{(j,i)}}{\max\{|\mathcal N_i|\sigma^{2}_{i},|\mathcal N_j|\sigma^{2}_{j}\}}$&$1-\sum_{k\neq j}A_j(k)$&
    \end{tabular}\label{hastings}
\end{eqnarray}
Table~\ref{table2} summarizes different fitness definitions corresponding to different combination rules in Table~\ref{table1}. Therefore, we can see that the existing algorithms can be summarized into our proposed graphical EGT framework with corresponding fitness definitions.

\subsection{Error-aware Distributed Adaptive Filtering Algorithm}

To illustrate our graphical EGT framework, as examples, we further design two distributed adaptive algorithms by choosing different fitness functions. As discussed in Section II, the existing distributed adaptive filtering algorithms either rely on the prior knowledge of network topology or the requirement of additional network statistics. All of them are not robust to a dynamic network, where a node location may change and the noise variance of each node may also vary with time. Considering these problems, we propose error-aware algorithms based on the intuition that neighbors with low mean-square-error (MSE) should be given more weight while neighbors with high MSE should be given less weight. The instantaneous error of node $i$, denoted by $\varrho_i$, can be calculated by
\begin{equation}
\varrho_{i,t}=\left|d_i(t)-\bm u_{i,t}\bm w_{i,t-1}\right|^2,
\end{equation}
where only local data $\{d_i(t), \bm u_{i,t}\}$ are used. The approximated MSE of node $i$, denoted by $\beta_{i}$, can be estimated by following update rule in each time slot,
\begin{equation}
\beta_{i,t}=(1-\nu_{i,t})\beta_{i,t-1}+\nu_{i,t}\varrho_{i,t},
\end{equation}
where $\nu_{i,t}$ is a positive parameter. We assume that nodes can exchange their instantaneous MSE information with neighbors. Based on the estimated MSE, we design two kinds of fitness: exponential form and power form as follows:
\begin{eqnarray}
\mbox{Power:}\!\!\!&&\!\!\!f_i=\beta_i^{-\lambda},\label{proposedfit2}\\
\mbox{Exponential:}\!\!\!&&\!\!\!f_i=e^{-\lambda \beta_i},\label{proposedfit}
\end{eqnarray}
where $\lambda$ is a positive coefficient. Note that the fitness defined in (\ref{proposedfit2}) and (\ref{proposedfit}) are just two examples of our proposed framework, while many other forms of fitness can be considered, e.g., $f_i=\log(\lambda\beta_i^{-1})$. Using the IM update rule, we have
\begin{eqnarray}
\bm w_{i,t+1}\!\!\!&=&\!\!\!\sum\limits_{j\in\mathcal{N}_i}\frac{\beta_{j,t}^{-\lambda}}{\sum_{k\in \mathcal N_i} \beta_{k,t}^{-\lambda}}F(\bm w_{j,t}),\label{proposed2}\\
\bm w_{i,t+1}\!\!\!&=&\!\!\!\sum\limits_{j\in\mathcal{N}_i}\frac{e^{-\lambda\beta_{j,t}}}{\sum_{k\in \mathcal N_i} e^{-\lambda\beta_{k,t}}}F(\bm w_{j,t}).\label{proposed}
\end{eqnarray}
From (\ref{proposed2}) and (\ref{proposed}), we can see that the proposed algorithms do not directly depend on any network topology information. Moreover, they can also adapt to a dynamic environment when the noise variance of nodes are unknown or suddenly change, since the weights can be immediately adjusted accordingly. In \cite{example}, a similar algorithm was also proposed based on the instantaneous MSE information, which is a special case of our error-aware algorithm with power form of $\lambda=2$. Note that the deterministic coefficients are adopted when implementing (\ref{proposed2}) and (\ref{proposed}), instead of using random combining efficient with some probability. However, the algorithm can also be implemented using a random selection with probabilities. There will be no performance loss since the expected outcome is the same, but the efficiency (convergence speed) will be lower. In Section V, we will verify the performance of the proposed algorithm through simulation.

\section{Diffusion Analysis}

In a distributed adaptive filter network, there are nodes with good signals, i.e., lower noise variance, as well as nodes with poor signals. The principal objective of distributed adaptive filtering algorithms is to stimulate the diffusion of good signals to the whole network to enhance the network performances. In this section, we will use the EGT to analyze such a dynamic diffusion process and derive the close-form expression for the diffusion probability. In the following diffusion analysis, we assume that all nodes have the same regressor statistics $\bf R_u$, but different noise statistics.

\begin{figure}[!t]
  \centerline{\epsfig{figure=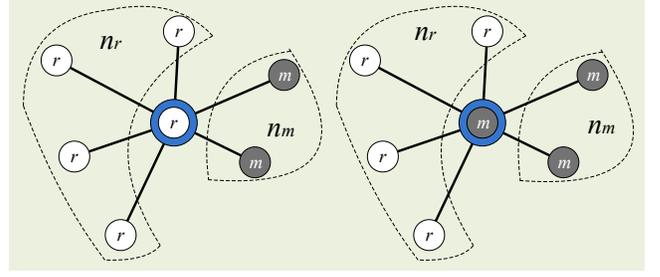,width=8.5cm}}
  \caption{Graphical evolutionary game model.}\label{fig4}
\end{figure}

In a graphical evolutionary game, the structured population are either residents or mutants. An important concept is the fixation probability, which represents the probability that the mutant will eventually overtake the whole population \cite{fixation}. Let us consider a local adaptive filter network as shown in Fig.\,\ref{fig4}, where the hollow points denote common nodes, i.e., nodes with common noise variance $\sigma_r^2$; and the solid points denote good nodes, i.e., nodes with a lower noise variance $\sigma_m^2$. $\sigma^2_r$ and $\sigma^2_m$ satisfy that $\sigma^2_r>>\sigma^2_m$. Here, we adopt the binary signal model to better reveal the diffusion process of good signals. If we regard the common nodes as residents and the good nodes as mutants, the concept of fixation probability in EGT can be applied to analyze the diffusion of good signals in the network. According to the definition of fixation probability, we define the diffusion probability in a distributed filter network as the probability that a good signal can be adopted by all nodes to update parameters in the network.

\subsection{Strategies and Utility Matrix}
As shown in Fig.\,\ref{fig4}, for the node at the center, its neighbors include both common nodes and good nodes. When the center node updates its parameter $\bm w_i$, it has the following two possible strategies:
\begin{eqnarray}
\left\{\begin{array}{ll}
\bm S_r,&\mbox{using information from common nodes}, \vspace{2mm}\\
\bm S_m,&\mbox{using information from good nodes}.
\end{array}\right.\label{strategy}
\end{eqnarray}
In such a case, we can define the utility matrix as follow:
\begin{eqnarray}
    \begin{tabular}{ccccc}
    && $\bm S_r$ &$\bm S_m$&\\
    \!\!\!\!$\bm S_r$ \!\!\!\!\!\!\!\!& \multirow{2}{0.01cm}{\bigg(}& $\pi^{-1}(\sigma_r,\sigma_r)$&$\pi^{-1}(\sigma_m,\sigma_r)$&\!\!\!\!\!\!\!\!\!\!\multirow{2}{0.01mm}{\bigg)\!\!\!\!}\\
    \!\!\!\!$\bm S_m$ \!\!\!\!\!\!\!\!& & $\pi^{-1}(\sigma_r,\sigma_m)$&$\pi^{-1}(\sigma_m,\sigma_m)$&
    \end{tabular}\hspace{-5mm}
    \begin{tabular}{ccccc}
    \!\!\!\!\multirow{4}{0.01cm}{=}\!\!\!\!\!\!\!\!\!&&&&\\
    &\multirow{2}{0.01cm}{\bigg(}& $u_1$&$u_2$&\!\!\!\!\!\!\!\!\!\!\multirow{2}{0.01mm}{\bigg)}\\
    &&$u_3$&$u_4$&
    \end{tabular}\!\!\!\!,\label{payoff}
\end{eqnarray}
where $\pi(x,y)$ represents the steady EMSE of node with noise variance $x^2$ using information from node with noise variance $y^2$. For example, $\pi(\sigma_r,\sigma_m)$ is the steady EMSE of node with noise variance $\sigma^2_r$ adopting strategy $\bm S_m$, i.e., updating its $\bm w$ using information from node with noise variance $\sigma^2_m$ which in turn adopts strategy $\bm S_r$. In our diffusion analysis, we assume that only two players are interacting with each other at one time instant, i.e., there are two nodes exchanging and combining information with each other at one time instant. In such a case, the payoff matrix is two-user case. Note that a node chooses one specific neighbor with some probability, which is equivalent to the weight that the node gives to that neighbor.

Since the steady EMSE $\pi(x,y)$ in the utility matrix is determined by the information combining rule, there is no general expressions for $\pi(x,y)$. Nevertheless, by intuition, we know that the steady EMSE of node with variance $\sigma^2_r$ should be larger than that of node with variance $\sigma^2_m$ since $\sigma^2_r>>\sigma^2_m$, and adopting strategy $\bm S_m$ should be more beneficial than adopting strategy $\bm S_r$ since the node can obtain better information from others, i.e., $\pi(\sigma_r,\sigma_r)>\pi(\sigma_r,\sigma_m)>\pi(\sigma_m,\sigma_r)>\pi(\sigma_m,\sigma_m)$. Therefore, we assume that the utility matrix defined in (\ref{payoff}) has the quality as follow
\begin{equation}
 u_1< u_3< u_2<u_4.\label{payoffq}
\end{equation}
Here, we use an example in \cite{two} to have a close-form expression for $\pi(x,y)$ to illustrate and verify this intuition. According to \cite{two}, with sufficiently small step size $\mu$, the optimal $\pi(x,y)$ can be calculated by
\begin{equation}
\pi(x,y)=c_1\sigma^2_{1}+c_2\frac{x^4}{\sigma_{2}^2},\label{pi}
\end{equation}
\begin{equation}
\left\{\begin{array}{l}
c_1=\frac{\mu\mbox{\scriptsize Tr}(\bm R_u)}{4},\quad\ c_2=\frac{\mu^2||\zeta||^2}{2}, \vspace{2mm}\\
\sigma^2_{1}=\frac{2x^2y^2}{x^2+y^2}, \quad \sigma^2_{2}=\frac{x^2y^2}{2},
\end{array}\right.\label{pi2}
\end{equation}
where $\zeta=\mbox{col}\{\zeta_1,...,\zeta_N\}$ consists of the eigenvalues of $\bm R_u$ (recall that $\bm R_u$ is the covariance matrix of the observed regression data $\bm u_t$). According to (\ref{pi}) and (\ref{pi2}), we have
\begin{eqnarray}
\pi(\sigma_r,\sigma_r)&\!\!\!=\!\!\!&c_1\sigma^2_{r}+2c_2,\label{pirr}\\
\pi(\sigma_r,\sigma_m)&\!\!\!=\!\!\!&c_1\frac{2\sigma^2_m\sigma^2_r}{\sigma^2_m+\sigma^2_r}+2c_2\frac{\sigma^2_r}{\sigma_{m}^2},\label{pirm}\\
\pi(\sigma_m,\sigma_r)&\!\!\!=\!\!\!&c_1\frac{2\sigma^2_m\sigma^2_r}{\sigma^2_m+\sigma^2_r}+2c_2\frac{\sigma^2_m}{\sigma_{r}^2},\label{pimr}\\
\pi(\sigma_m,\sigma_m)&\!\!\!=\!\!\!&c_1\sigma^2_{m}+2c_2.\label{pimm}
\end{eqnarray}
Suppose $\sigma_m^2=\tau\sigma_r^2$, through comparing (\ref{pirr}-\ref{pimm}), we can derive the condition for $\pi(\sigma_r,\sigma_r)>\pi(\sigma_r,\sigma_m)>\pi(\sigma_m,\sigma_r)>\pi(\sigma_m,\sigma_m)$ as follows
\begin{equation}
\mu<\frac{\tau\mbox{Tr}(\mathbf R_u)\sigma_r^2}{4(1+\tau)||\zeta||^2}.\label{coditionmu}
\end{equation}
According to \cite{two}, the derivation of optimal $\pi(x,y)$ in (\ref{pi}) and (\ref{pi2}) is based on the assumption that $\mu$ is sufficiently small. Therefore, the condition of $\mu$ in (\ref{coditionmu}) holds. In such a case, we can conclude that $\pi(\sigma_r,\sigma_r)>\pi(\sigma_r,\sigma_m)>\pi(\sigma_m,\sigma_r)>\pi(\sigma_m,\sigma_m)$, which implies that
$u_1< u_3< u_2<u_4$.

In the following, we will analyze the diffusion process of strategy $\bm S_m$, i.e., the ability of good signals diffusing over the whole network. We consider an adaptive filter network based on a homogenous graph with general degree $n$ and adopt the IM update rule for the parameter update \cite{core}. Let $p_r$ and $p_m$ denote the percentages of nodes using strategies $\bm S_r$ and $\bm S_m$ in the population, respectively. Let $p_{rr}$, $p_{rm}$, $p_{mr}$ and $p_{mm}$ denote the percentages of edge, where $p_{rm}$ means the percentage of edge on which both nodes use strategy $\bm S_r$ and $\bm S_m$. Let $q_{m|r}$ denote the conditional probability of a node using strategy $\bm S_m$ given that the adjacent node is using strategy $\bm S_r$, similar we have $q_{r|r}$, $q_{r|m}$ and $q_{m|m}$. In such a case, we have
\begin{gather}
p_r+p_m=1,\label{eqset1}\quad
q_{r|X}+q_{m|X}=1,\\
p_{XY}=p_Y\cdot q_{X|Y},\quad
p_{rm}=p_{mr}\label{eqset4},
\end{gather}
where $X$ and $Y$ are either $r$ or $m$. The equations (\ref{eqset1}-\ref{eqset4}) imply that the state of the whole network can be described by only two variables, $p_m$ and $q_{m|m}$. In the following, we will calculate the dynamics of $p_m$ and $q_{m|m}$ under the IM update rule.

\subsection{Dynamics of $p_m$ and $q_{m|m}$}

In order to derive the diffusion probability, we first need to analyze the diffusion process of the system. As discussed in the previous subsection, the system dynamics under IM update rule can be represented by parameters $p_m$ and $q_{m|m}$. Thus, in this subsection, we will first analyze the dynamics of $p_m$ and $q_{m|m}$ to understand the dynamic diffusion process of the adaptive network. According to the IM update rule, a node using strategy $\bm S_r$ is selected for imitation with probability $p_r$. As shown in the left part of Fig.\,\ref{fig4}, among its $n$ neighbors (not including itself), there are $n_r$ nodes using strategy $\bm S_r$ and $n_m$ nodes using strategy $\bm S_m$, respectively, where $n_r+n_m=n$. The percentage of such a configuration is $\binom{n}{n_m}q_{m|r}^{n_m}q_{r|r}^{n_r}$. In such a case, the fitness of this node is
\begin{equation}
f_0=(1-\alpha)+\alpha(n_ru_1+n_mu_2),\label{f0}
\end{equation}
where the baseline fitness is normalized as $1$. We can see that (\ref{f0}) includes the normalized baseline fitness and also the fitness from utility, which is the standard definition of fitness used in the EGT filed, as shown in (\ref{fitness}). Among those $n$ neighbors, the fitness of node using strategy $\bm S_m$ is
\begin{equation}
f_m=(1-\alpha)+\alpha\Big(\big[(n-1)q_{r|m}+1\big] u_3+(n-1)q_{m|m} u_4\Big),
\end{equation}
and the fitness of node using strategy $\bm S_r$ is
\begin{equation}
f_r=(1-\alpha)+\alpha\Big(\big[(n-1)q_{r|r}+1\big] u_1+(n-1)q_{m|r} u_2\Big).
\end{equation}
In such a case, the probability that the node using strategy $\bm S_r$ is replaced by $\bm S_m$ is
\begin{equation}
P_{r\rightarrow m}=\frac{n_mf_m}{n_mf_m+n_rf_r+f_0}.
\end{equation}
Therefore, the percentage of nodes using strategy $\bm S_m$, $p_m$, increases by $1/N$ with probability
\begin{eqnarray}
\mbox{Prob}\Big(\Delta p_m=\frac{1}{N}\Big)\!\!\!&=&\!\!\!p_r\sum\limits_{n_r+n_m=n}\binom{n}{n_m}q_{m|r}^{n_m}q_{r|r}^{n_r}\nonumber\\
\!\!\!&&\!\!\!\cdot\frac{n_mf_m}{n_mf_m+n_rf_r+f_0}.\label{pmin}
\end{eqnarray}
Meanwhile, the edges that both nodes use strategy $\bm S_m$ increase by $n_m$, thus, we have
\begin{eqnarray}
\mbox{Prob}\Big(\Delta p_{mm}=\frac{2n_m}{nN}\Big)\!\!\!&=&\!\!\!p_r\binom{n}{n_m}q_{m|r}^{n_m}q_{r|r}^{n_r}\nonumber\\
\!\!\!&&\!\!\!\cdot\frac{n_mf_m}{n_mf_m+n_rf_r+f_0}.\label{pmmin}
\end{eqnarray}

Similar analysis can be applied to the node using strategy $\bm S_m$. According to the IM update rule, a node using strategy $\bm S_m$ is selected for imitation with probability $p_m$. As shown in the right part of Fig.\,\ref{fig4}, we also assume that there are $n_r$ nodes using strategy $\bm S_r$ and $n_m$ nodes using strategy $\bm S_m$ among its $n$ neighbors. The percentage of such a phenomenon is $\binom{n}{n_m} q_{m|m}^{n_m}q_{r|m}^{n_r}$. Thus, the fitness of this node is
\begin{equation}
g_0=(1-\alpha)+\alpha(n_ru_2+n_mu_3).
\end{equation}
Among those $n$ neighbors, the fitness of node using strategy $\bm S_m$ is
\begin{equation}
g_m=(1-\alpha)+\alpha\Big((n-1)q_{r|m} u_3+\big[(n-1)q_{m|m}+1\big] u_4\Big),
\end{equation}
and the fitness of node using strategy $\bm S_r$ is
\begin{equation}
g_r=(1-\alpha)+\alpha\Big((n-1)q_{r|r} u_1+\big[(n-1)q_{m|r}+1\big] u_2\Big).
\end{equation}
In such a case, the probability that the node using strategy $\bm S_m$ is replaced by $\bm S_r$ is
\begin{equation}
P_{m\rightarrow r}=\frac{n_rg_r}{n_mg_m+n_rg_r+g_0}.
\end{equation}
Therefore, the percentage of nodes using strategy $\bm S_m$, $p_m$, decreases by $1/N$ with probability
\begin{eqnarray}
\mbox{Prob}\Big(\Delta p_m=-\frac{1}{N}\Big)\!\!\!&=&\!\!\!p_m\sum\limits_{n_r+n_m=n}\binom{n}{n_m}q_{m|m}^{n_m}q_{r|m}^{n_r}\nonumber\\
\!\!\!&&\!\!\!\cdot\frac{n_rg_r}{n_mg_m+n_rg_r+g_0}.\label{pmde}
\end{eqnarray}
Meanwhile, the edges that both nodes use strategy $\bm S_m$ decrease by $n_m$, thus, we have
\begin{eqnarray}
\mbox{Prob}\Big(\Delta p_{mm}=-\frac{2n_m}{nN}\Big)\!\!\!&=&\!\!\!p_m\binom{n}{n_m}q_{m|m}^{n_m}q_{r|m}^{n_r}\nonumber\\
\!\!\!&&\!\!\!\cdot\frac{n_rg_r}{n_mg_m+n_rg_r+g_0}.\label{pmmde}
\end{eqnarray}

Combining (\ref{pmin}) and (\ref{pmde}), we have the dynamic of $p_m$ as
\begin{eqnarray}
\!\!\!\!\!\!\!\!\!\!\!\!\!\!&&\!\!\!\!\dot p_m=\frac{1}{N}\mbox{Prob}\Big(\Delta p_m=\frac{1}{N}\Big)-\frac{1}{N}\mbox{Prob}\Big(\Delta p_m=-\frac{1}{N}\Big)\nonumber\\
\!\!\!\!\!\!\!\!\!\!\!\!\!\!&=&\!\!\!\!\!\frac{\alpha n(n-1)p_{rm}}{N(n+1)^2}(\gamma_1u_1+\gamma_2u_2+\gamma_3u_3+\gamma_4u_4)\!+\!O(\alpha^2),\label{dpm}
\end{eqnarray}
where the second equality is according to Taylor's Theorem and weak selection assumption with $\alpha$ goes to zero \cite{weak0}, and the parameters $\gamma_1$, $\gamma_2$, $\gamma_3$ and $\gamma_4$ are given as follows:
\begin{eqnarray}
\!\!\!\!\!\!\!\gamma_1\!\!\!\!\!&=&\!\!\!\!-q_{r|r}[(n-1)(q_{r|r}+q_{m|m})+3],\\
\!\!\!\!\!\!\!\gamma_2\!\!\!\!\!&=&\!\!\!\!-q_{m|m}-q_{m|r}[(n-1)(q_{r|r}+q_{m|m})+2]-\frac{2}{n\!-\!1},\\
\!\!\!\!\!\!\!\gamma_3\!\!\!\!\!&=&\!\!\!\!q_{r|r}+q_{r|m}[(n-1)(q_{r|r}+q_{m|m})+2]+\frac{2}{n\!-\!1},\\
\!\!\!\!\!\!\!\gamma_4\!\!\!\!\!&=&\!\!\!\!q_{m|m}[(n-1)(q_{r|r}+q_{m|m})+3].
\end{eqnarray}
In (\ref{dpm}), the dot notation $\dot p_m $ represents the dynamic of $p_m$, i.e., the variation of $p_m$ within a tiny period of time. In such a case, the utility obtained from the interactions is considered as limited contribution to the overall fitness of each player. On one hand, the results derived from weak selection often remain as valid approximations for larger selection strength \cite{weak}. On the other hand, the weak selection limit has a long tradition in theoretical biology \cite{weak1}. Moreover, the weak selection assumption can help achieve a close-form analysis of diffusion process and better reveal how the strategy diffuses over the network.
Similarly, by combining (\ref{pmmin}) and (\ref{pmmde}), we have the dynamics of $p_{mm}$ as
\begin{eqnarray}
\!\!\!\!\!\dot p_{mm}\!\!\!\!&=&\!\!\!\!\sum\limits_{n_m=0}^n\frac{2n_m}{nN}\mbox{Prob}\Big(\Delta p_{mm}=\frac{2n_m}{nN}\Big)\nonumber\\
\!\!\!\!&&\!\!\!\!-\sum\limits_{n_m=0}^n\frac{2n_m}{nN}\mbox{Prob}\Big(\Delta p_{mm}=-\frac{2n_m}{nN}\Big)\nonumber\\
\!\!\!\!&=&\!\!\!\!\frac{2p_{rm}}{(n+1)N}\Big(1+(n-1)(q_{m|r}\!-\!q_{m|m})\Big)\!+\!O(\alpha).
\end{eqnarray}
Besides, we can also have the dynamics of $q_{m|m}$ as
\begin{eqnarray}
\!\!\!\!\!\!\!\!\!\!\!\!\!\!\!\!\!\!\!\!&&\dot q_{m|m}=\frac{d}{dt}\Big(\frac{p_{mm}}{p_m}\Big)\nonumber\\
\!\!\!\!\!\!\!\!\!\!\!\!\!\!\!\!\!\!\!\!&&=\frac{2}{(n+1)N}\frac{p_{rm}}{p_m}\Big(1+(n-1)(q_{m|r}-q_{m|m})\Big)+O(\alpha).\label{dqmm}
\end{eqnarray}

\subsection{Diffusion Probability Analysis}
The dynamic equation of $p_m$ in (\ref{dpm}) reflects the the dynamic of nodes updating $\bm w$ using information from good nodes, i.e., the diffusion status of good signals in the network. A positive $\dot p_m$ means that good signals are diffusing over the network, while a negative $\dot p_m$ means that good signals have not been well adopted. The diffusion probability of good signals is closely related to the noise variance of good nodes $\sigma_m$. Intuitively, the lower $\sigma_m$, the higher probability that good signals can spread the whole network. In this subsection, we will analyze the close-form expression for the diffusion probability.

As discussed at the beginning of Section IV, the state of whole network can be described by only $p_m$ and $q_{m|m}$. In such a case, (\ref{dpm}) and (\ref{dqmm}) can be re-written as functions of $p_m$ and $q_{m|m}$
\begin{eqnarray}
\dot p_m\!\!\!&=&\!\!\!\alpha\cdot G_1(p_m,q_{m|m})+O(\alpha^2),\label{dpm2} \\
\dot q_{m|m}\!\!\!&=&\!\!\!G_2(p_m,q_{m|m})+O(\alpha)\label{dqmm2}.
\end{eqnarray}
From (\ref{dpm2}) and (\ref{dqmm2}), we can see that $q_{m|m}$ converges to equilibrium in a much faster rate than $p_m$ under the assumption of weak selection. At the steady state of $q_{m|m}$, i.e., $\dot q_{m|m}=0$, we have
\begin{equation}
q_{m|m}-q_{m|r}=\frac{1}{n-1}.\label{steady}
\end{equation}
In such a case, the dynamic network will rapidly converge onto the slow manifold, defined by $G_2(p_m,q_{m|m})=0$. Therefore, we can assume that (\ref{steady}) holds in the whole convergence process of $p_m$. According to (\ref{eqset1})-(\ref{eqset4}) and (\ref{steady}), we have
\begin{eqnarray}
q_{m|m}\!\!\!&=&\!\!\!p_m+\frac{1}{n-1}(1-p_m),\\
q_{m|r}\!\!\!&=&\!\!\!\frac{n-2}{n-1}p_m,\\
q_{r|m}\!\!\!&=&\!\!\!\frac{n-2}{n-1}(1-p_m),\\
q_{r|r}\!\!\!&=&\!\!\!1-\frac{n-2}{n-1}p_m\label{steady2}.
\end{eqnarray}
Therefore, the diffusion process can be characterized by only $p_m$. Thus, we can focus on the dynamics of $p_m$ to derive the diffusion probability, which is given by following \emph{Theorem 1}.

\emph{Theorem 1:} In a distributed adaptive filter network which can be characterized by a $N$-node regular graph with degree $n$, suppose there are common nodes with noise variance $\sigma_r$ and good nodes with noise variance $\sigma_m$, where each common node has connection edge with only one good node. If each node updates its parameter $\bm w$ using the IM update rule, the diffusion probability of the good signal can be approximated by
\begin{equation}
P_{\mbox{\scriptsize diff}}=\frac{1}{n+1}+\frac{\alpha nN}{6(n+1)^3}(\xi_1u_1+\xi_2u_2+\xi_3u_3+\xi_4u_4),\label{difc}
\end{equation}
where the parameters $\xi_1$, $\xi_2$, $\xi_3$ and $\xi_4$ are as follows:
\begin{gather}
\xi_1=-2n^2-5n+3,\quad
\xi_2=-n^2-n-3,\\
\xi_3=2n^2+2n-3,\quad
\xi_4=n^2+4n+3.
\end{gather}
\begin{proof}
See Appendix.
\end{proof}

Using \emph{Theorem 1}, we can calculate the diffusion probability of the good signals over the network, which can be used to evaluate the performance of an adaptive filter network. Similarly, the diffusion dynamics and probabilities under BD and DB update rules can also be derived using the same analysis. The following theorem shows an interesting result, which is based on an important theorem in \cite{gegreview}, stating that evolutionary dynamics under BD, DB, and IM are equivalent for undirected regular graphs.

\emph{Theorem 2:} In a distributed adaptive filter network which can be characterized by a $N$-node regular graph with degree $n$, suppose there are common nodes with noise variance $\sigma_r$ and good nodes with noise variance $\sigma_m$, where each common node has connection edge with only one good node. If each node updates its parameter $\bm w$ using the IM update rule, the diffusion probabilities of good signals under BD and DB update rules are same with that under the IM update rule.

\section{Evolutionarily Stable Strategy}

In the last section, we have analyzed the information diffusion process in an adaptive network under the IM update rule, and derived the diffusion probability of strategy $\bm S_m$ that using information from good nodes. On the other hand, considering that if the whole network has already chosen to adopt this favorable strategy $\bm S_m$, is the current state a stable network state, even though a small fraction of nodes adopt the other strategy $\bm S_r$? In the following, we will answer these questions using the concept of evolutionarily stable strategy (ESS) in evolutionary game theory. As discussed in Section III-A, the ESS ensures that one strategy is resistant against invasion of another strategy \cite{ess}. In our system model, it is obvious that $\bm S_m$, i.e., using information from good nodes, is the favorable strategy and a desired ESS in the network. In this section, we will check whether strategy $\bm S_m$ is evolutionarily stable.

\subsection{ESS in Complete Graphs}

We first discuss whether strategy $\bm S_m$ is an ESS in complete graphs, which is shown by the following theorem.

\emph{Theorem 3:} In a distributed adaptive filter network that can be characterized by complete graphs, strategy $\bm S_m$ is always an ESS strategy.
\begin{proof}
In a complete graph, each node meets every other node equally likely. In such a case, according to the utility matrix in (\ref{payoff}), the average utilities of using strategies $\bm S_r$ and $\bm S_m$ are given by
\begin{eqnarray}
U_r\!\!\!&=&\!\!\!p_ru_1+p_mu_2,\\
U_m\!\!\!&=&\!\!\!p_ru_3+p_mu_4,
\end{eqnarray}
where $p_r$ and $p_m$ are the percentages of population using strategies $\bm S_r$ and $\bm S_m$, respectively. Consider the scenario that the majority of the population adopt strategy $\bm S_m$, while a small fraction of the population adopt $\bm S_r$ which is considered as invasion, $p_r=\epsilon$. In such a case, according to the definition of ESS in (\ref{ess}), strategy $\bm S_m$ is evolutionary stable if $U_m>U_r$ for $(p_r,p_m)=(\epsilon,1-\epsilon)$, i.e.,
\begin{equation}
\epsilon(u_3-u_1)+(1-\epsilon)(u_4-u_2)>0.\label{essc}
\end{equation}
For $\epsilon\rightarrow 0$, the left hand side of (\ref{essc}) is positive if and only if
\begin{equation}
\mbox{``}u_4>u_2 \mbox{''}\quad \mbox{or}\quad \mbox{``}u_4=u_2 \ \mbox{and} \ u_3>u_1\mbox{''}.\label{esscondition}
\end{equation}
The (\ref{esscondition}) gives the sufficient evolutionary stable condition of strategy $\bm S_m$. In our system, we have $u_4>u_2>u_3>u_1$, which means that (\ref{esscondition}) always holds. Therefore, strategy $\bm S_m$ is always an ESS if the adaptive filter network is a complete graph.
\end{proof}

\begin{figure*}[!t]
\hspace{3mm}
\begin{minipage}[!t]{0.5\linewidth}
\centerline{\epsfig{figure=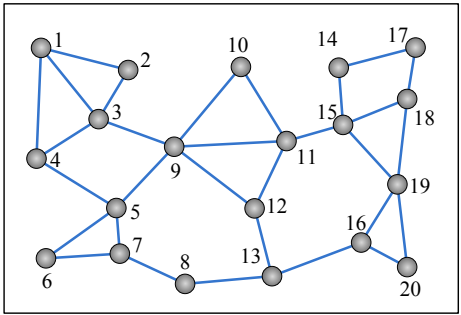,height=4.2cm}}
\end{minipage}
\hspace{-18mm}
\begin{minipage}[!t]{0.5\linewidth}
\centerline{\epsfig{figure=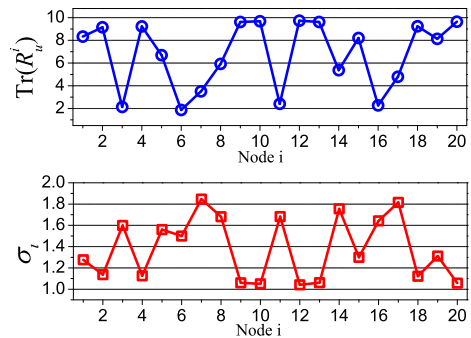,height=4.4cm}}
\end{minipage}
\caption{Network information for simulation, including network topology for $20$ nodes (left), trace of regressor covariance $\mbox{Tr}(\bm R_u)$ (right top) and noise variance $\sigma_i$ (right bottom).}\label{nettop}
\end{figure*}

\subsection{ESS in Incomplete Graphs}

Let us consider an adaptive filter network which can be characterized by an incomplete regular graph with degree $n$. The following theorem shows that strategy $\bm S_m$ is always an ESS in such an incomplete graph.

\emph{Theorem 4:} In a distributed adaptive filter network which can be characterized by a regular graph with degree $n$, strategy $\bm S_m$ is always an ESS strategy.
\begin{proof}
Using the pair approximation method \cite{rd}, the replicator dynamics of strategies $\bm S_m$ and $\bm S_r$ on a regular graph of degree $n$ can be approximated simply by
\begin{eqnarray}
\dot p_r\!\!\!&=&\!\!\!p_r(p_ru^\prime_1+p_mu^\prime_2-\phi),\\
\dot p_m\!\!\!&=&\!\!\!p_m(p_ru^\prime_3+p_mu^\prime_4-\phi),
\end{eqnarray}
where $\phi=p_rp_ru_1^\prime+p_rp_m(u_2^\prime+u_3^\prime)+p_mp_mu_4^\prime$ is the average utility, and $u^\prime_1$, $u^\prime_2$, $u^\prime_3$ and $u^\prime_4$ are given as follows:
\begin{equation}
\left\{\begin{array}{l}
u^\prime_1=u_1, \vspace{2mm}\\
u^\prime_2=u_2+u^\prime,\vspace{2mm}\\
u^\prime_3=u_3-u^\prime,\vspace{2mm}\\
u^\prime_4=u_4.
\end{array}\right.
\end{equation}
The parameter $u^\prime$ depends on the three update rules (IM, BD and DB), which is given by \cite{rd}
\begin{eqnarray}
\!\!\!\!\mbox{IM:}\!\!&&\!\!\!\!u^\prime=\frac{(n\!+3)u_1+u_2-u_3-(n\!+3)u_4}{(n+3)(n-2)}, \\
\!\!\!\!\mbox{BD:}\!\!&&\!\!\!\!u^\prime=\frac{(n\!+1)u_1+u_2-u_3-(n\!+1)u_4}{(n+1)(n-2)},\\
\!\!\!\!\mbox{DB:}\!\!&&\!\!\!\!u^\prime=\frac{u_1+u_2-u_3-u_4}{n-2}.
\end{eqnarray}
In such a case, the equivalent utility matrix is
\begin{eqnarray}
    \begin{tabular}{ccccc}
    && $\bm S_r$ &$\bm S_m$&\\
    \!\!\!\!$\bm S_r$ \!\!\!\!\!\!\!\!& \multirow{2}{0.01cm}{\bigg(}& $u_1$&$u_2+u^\prime$&\!\!\!\!\!\!\!\!\!\!\multirow{2}{0.01mm}{\bigg)\!\!\!\!}\\
    \!\!\!\!$\bm S_m$ \!\!\!\!\!\!\!\!& & $u_3-u^\prime$&$u_4$&\\
    &&&&
    \end{tabular}\!\!\!.\label{newpayoff}
\end{eqnarray}

According to (\ref{esscondition}), the evolutionary stable condition for strategy $\bm S_m$ is
\begin{equation}
u_4>u_2+u^\prime.\label{incon}
\end{equation}
Since $u_1<u_3<u_2<u_4$, we have $u^\prime<0$ for all three update rules. In such a case, (\ref{incon}) always holds, which means that strategy $\bm S_m$ is always an ESS strategy. This completes the proof of the theorem.
\end{proof}

\section{Simulation Results}

In this section, we develop simulations to compare the performances of different adaptive filtering algorithms, as well as to verify the derivation of information diffusion probability and the analysis of ESS.

\begin{figure}[!t]
  \centering
\begin{minipage}[b]{1\linewidth}
  \centering
  \centerline{\epsfig{figure=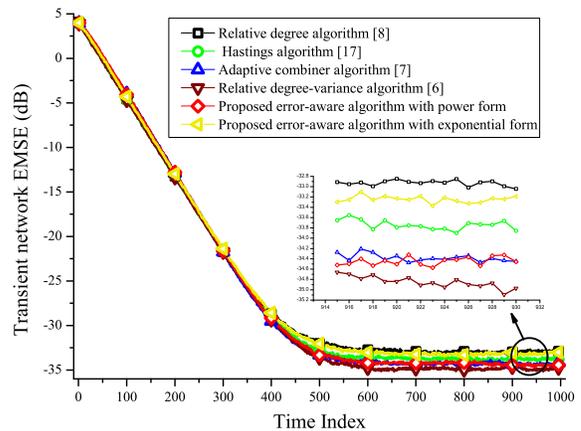,width=7.5cm}}
  \centerline{\scriptsize{(a) Network EMSE.}}\vspace{0.5cm}
\end{minipage}
\begin{minipage}[b]{1\linewidth}
  \centering
  \centerline{\epsfig{figure=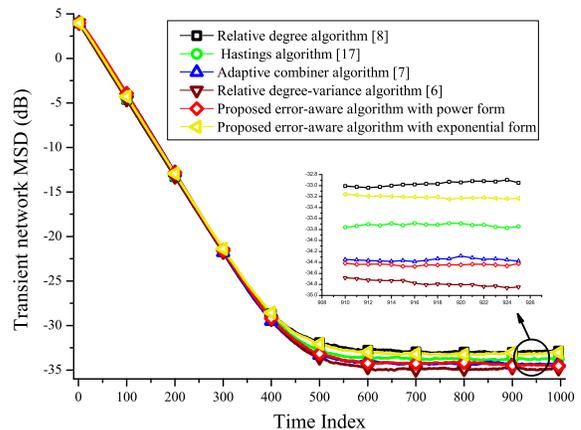,width=7.5cm}}
  \centerline{\scriptsize{(b) Network MSD.}}
\end{minipage}
\caption{Transient performances comparison with known noise variances.}\label{netmse}
\end{figure}

\subsection{Mean-square Performances}
The network topology used for simulation is shown in the left part of Fig.\,\ref{nettop}, where $20$ randomly nodes are randomly located. The signal and noise power information of each node are also shown in the right part of Fig.\,\ref{nettop}, respectively. In the simulation, we assume that the regressors with size $M=5$, are zero-mean Gaussian and independent in time and space. The unknown vector is set to be $\bm w^0=\mathds 1_5/\sqrt{2}$ and the step size of the LMS algorithm at each node $i$ is set as $\mu_i=0.01$. All the simulation results are averaged over $500$ independent runnings. All the performance comparisons are conducted among six different kinds of distributed adaptive filtering algorithms as follows:
\begin{itemize}
\item Relative degree algorithm \cite{rls};
\item Hastings algorithm \cite{two};
\item Adaptive combiner algorithm \cite{lms2};
\item Relative degree-variance algorithm \cite{lms};
\item Proposed error-aware algorithm with power form;
\item Proposed error-aware algorithm with exponential form.
\end{itemize}
Among these algorithms, the adaptive combiner algorithm \cite{lms2} and our proposed error-aware algorithm are based on dynamic combiners (weights), which are updated in each time slot. The difference is the updating rule, where the adaptive combiner algorithm in \cite{lms2} uses optimization and projection method, and our proposed algorithms use the approximated EMSE information.

\begin{figure}[!t]
  \centering
\begin{minipage}[b]{1\linewidth}
  \centering
  \centerline{\epsfig{figure=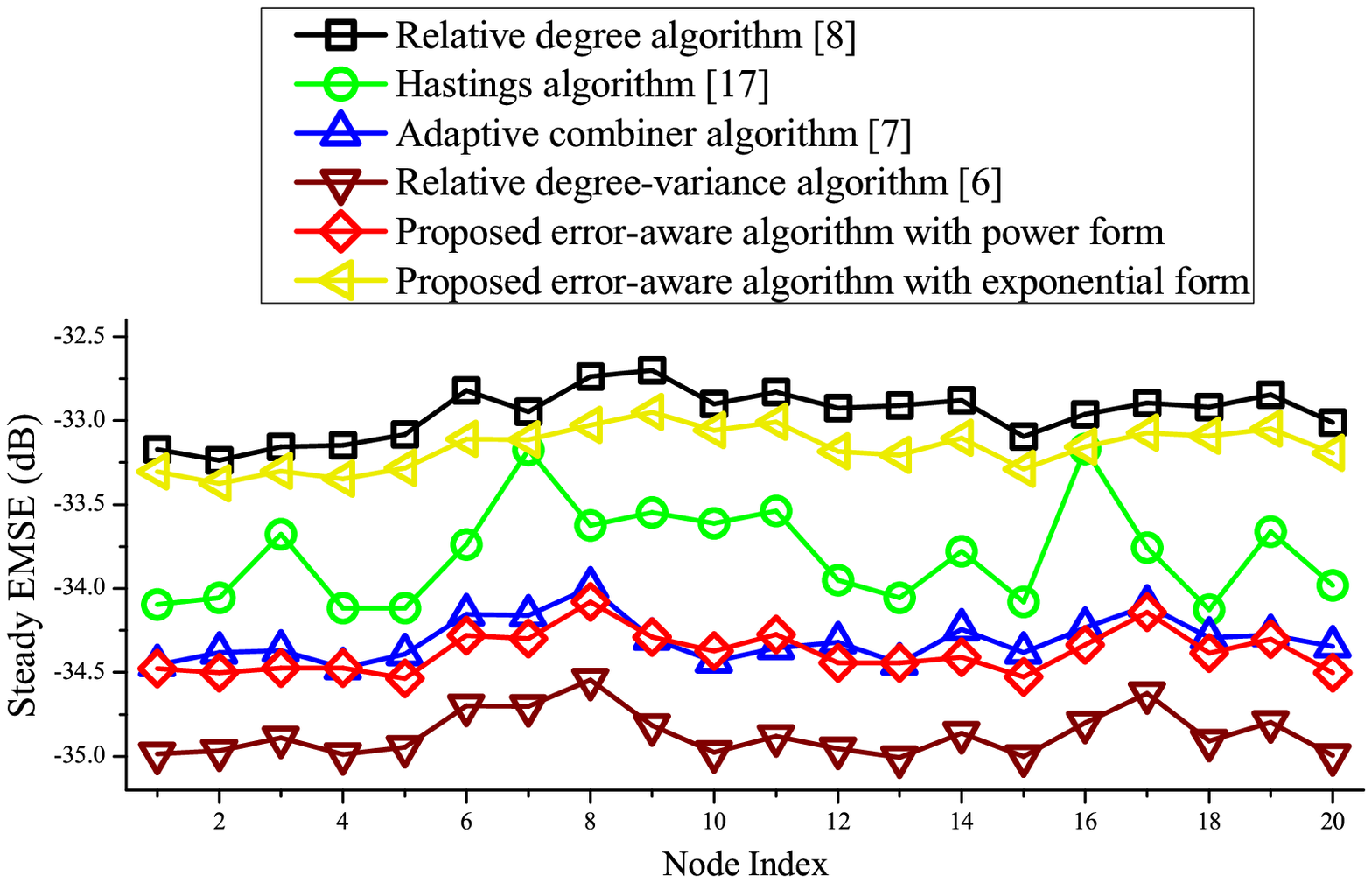,width=7.5cm}}
  \centerline{\scriptsize{(a) Node's EMSE.}}\vspace{0.1cm}
\end{minipage}
\begin{minipage}[b]{1\linewidth}
  \centering
  \centerline{\epsfig{figure=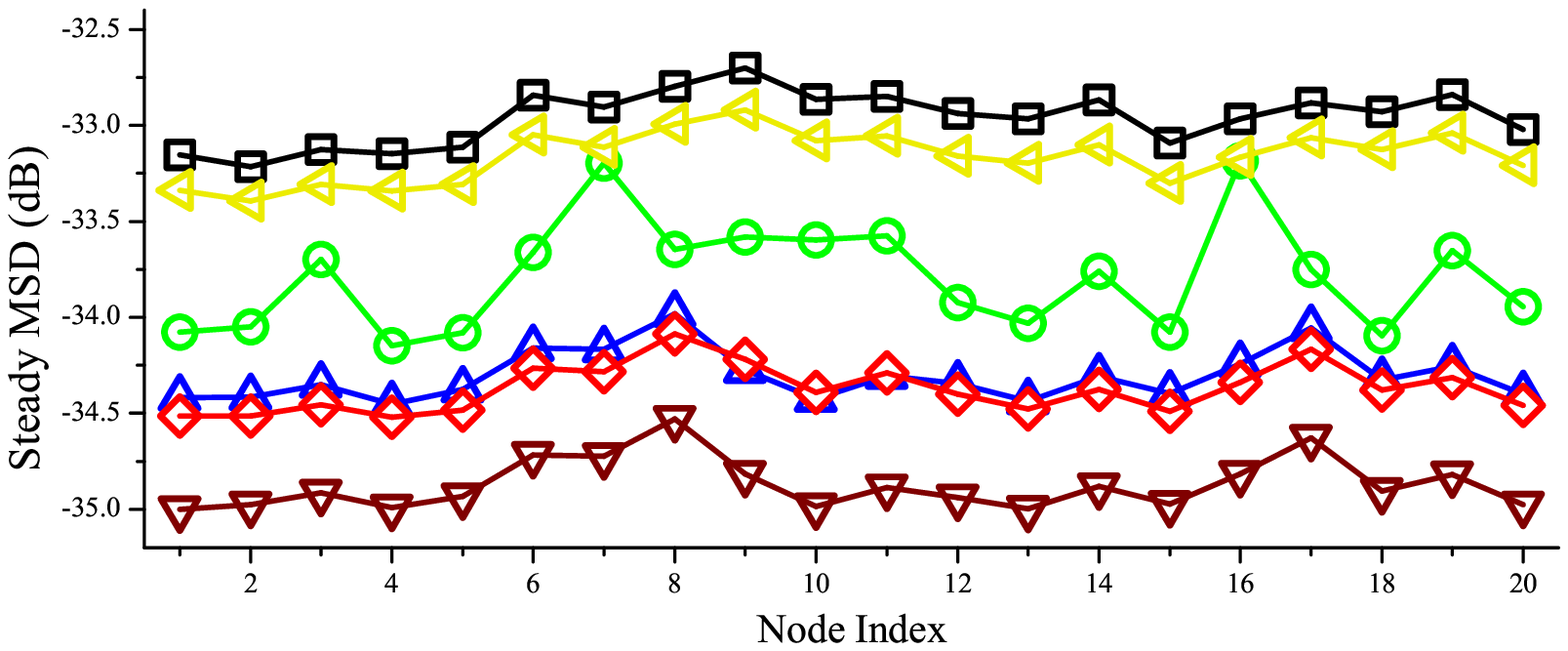,width=7.5cm}}
  \centerline{\scriptsize{(b) Node's MSD.}}
\end{minipage}
\caption{Steady performances comparison with known noise variances.}\label{steadynetmse}
\end{figure}

In the first comparison, we assume that the noise variance of each node is known by the Hastings and relative degree-variance algorithms. Fig.\,\ref{netmse} shows the transient network-performance comparison results among six kinds of algorithms in terms of EMSE and MSD. Under the similar convergence rate, we can see that the relative degree-variance algorithm performs the best. The proposed algorithm with exponential form performs better than the relative degree algorithm. With the power form fitness, the proposed algorithm can achieve similar performance, if not better than, compared with adaptive combiner algorithm, and both algorithms performs better than all other algorithms except the relative degree-variance algorithm. However, as discussed in Section 2, the relative degree-variance algorithm requires noise variance information of each node, while our proposed algorithm does not. Fig.\,\ref{steadynetmse} shows the corresponding steady-state performances of each node for six kinds of distributed adaptive filtering algorithms in terms of EMSE and MSD. Since the steady-state result is for each node, besides averaging over $500$ independent runnings, we average at each node over $100$ time slots after the convergence. We can see that the comparison results of steady-state performances are similar to those of the transient performances.

\begin{figure}[!t]
  \centering
\begin{minipage}[b]{1\linewidth}
  \centering
  \centerline{\epsfig{figure=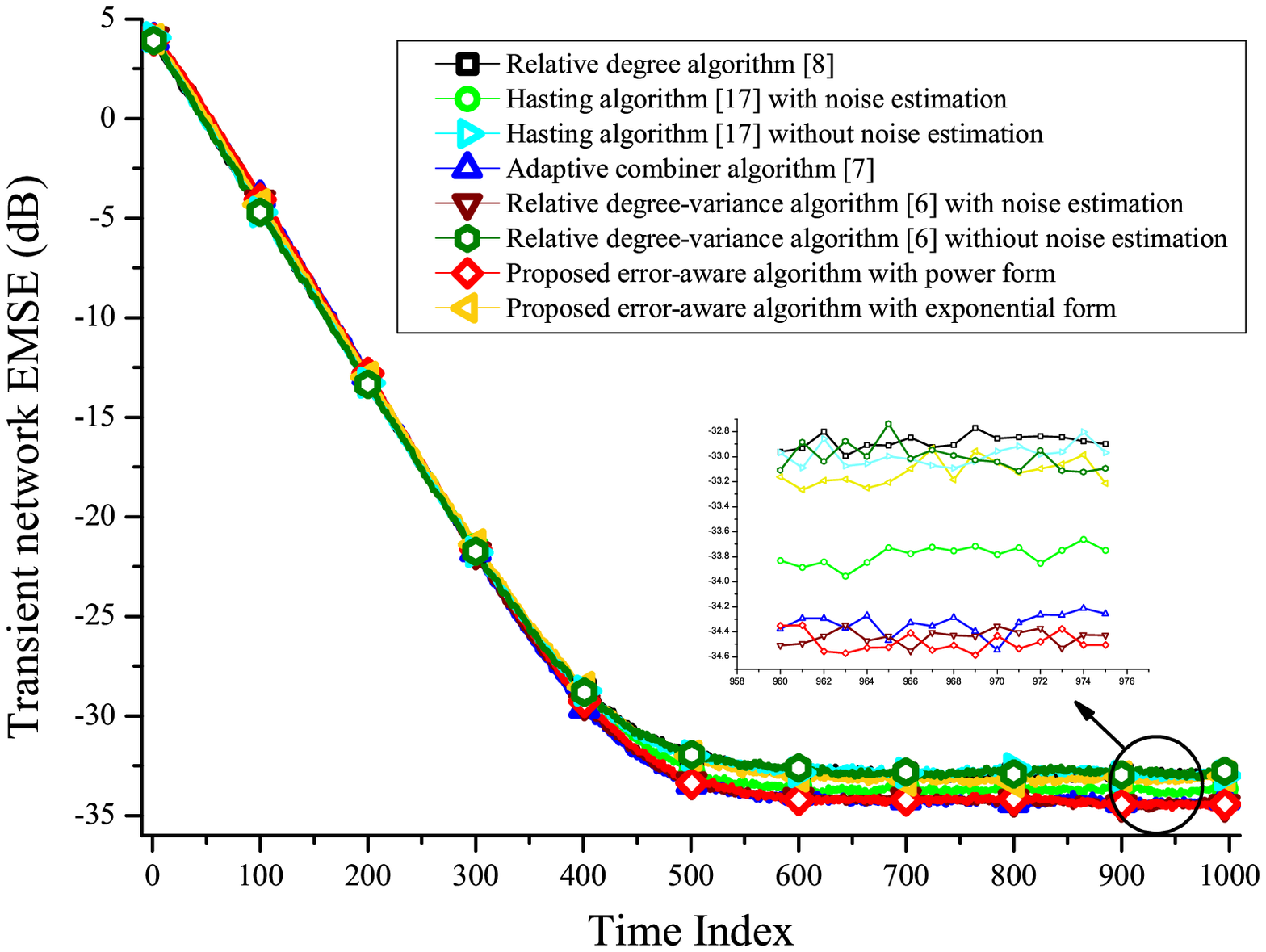,width=7.5cm}}
  \centerline{\scriptsize{(a) Network EMSE.}}\vspace{0.5cm}
\end{minipage}
\begin{minipage}[b]{1\linewidth}
  \centering
  \centerline{\epsfig{figure=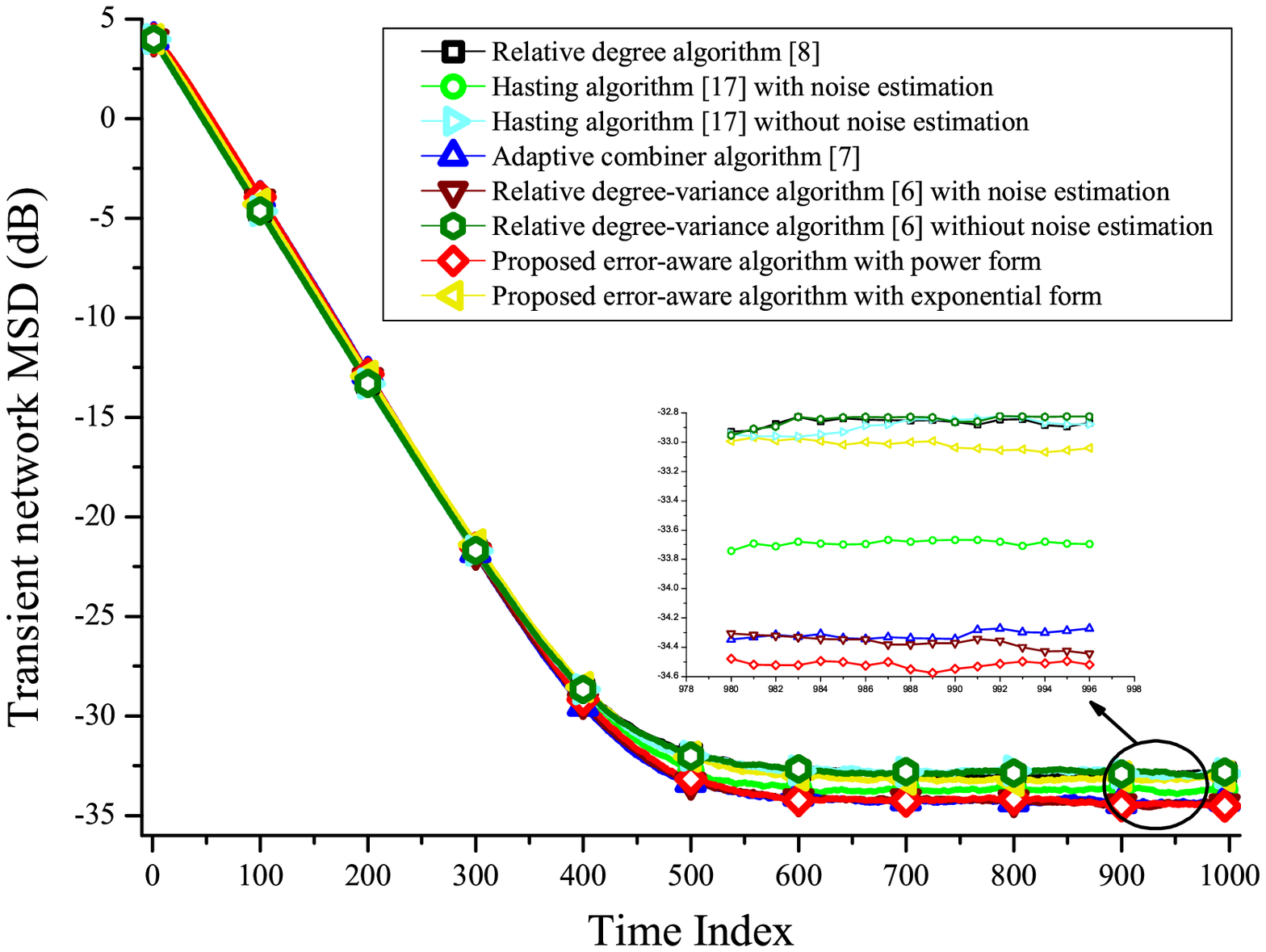,width=7.5cm}}
  \centerline{\scriptsize{(b) Network MSD.}}
\end{minipage}
\caption{Transient performances comparison with unknown noise variances.}\label{netmsev}
\end{figure}

\begin{figure}[!t]
  \centering
\begin{minipage}[b]{1\linewidth}
  \centering
  \centerline{\epsfig{figure=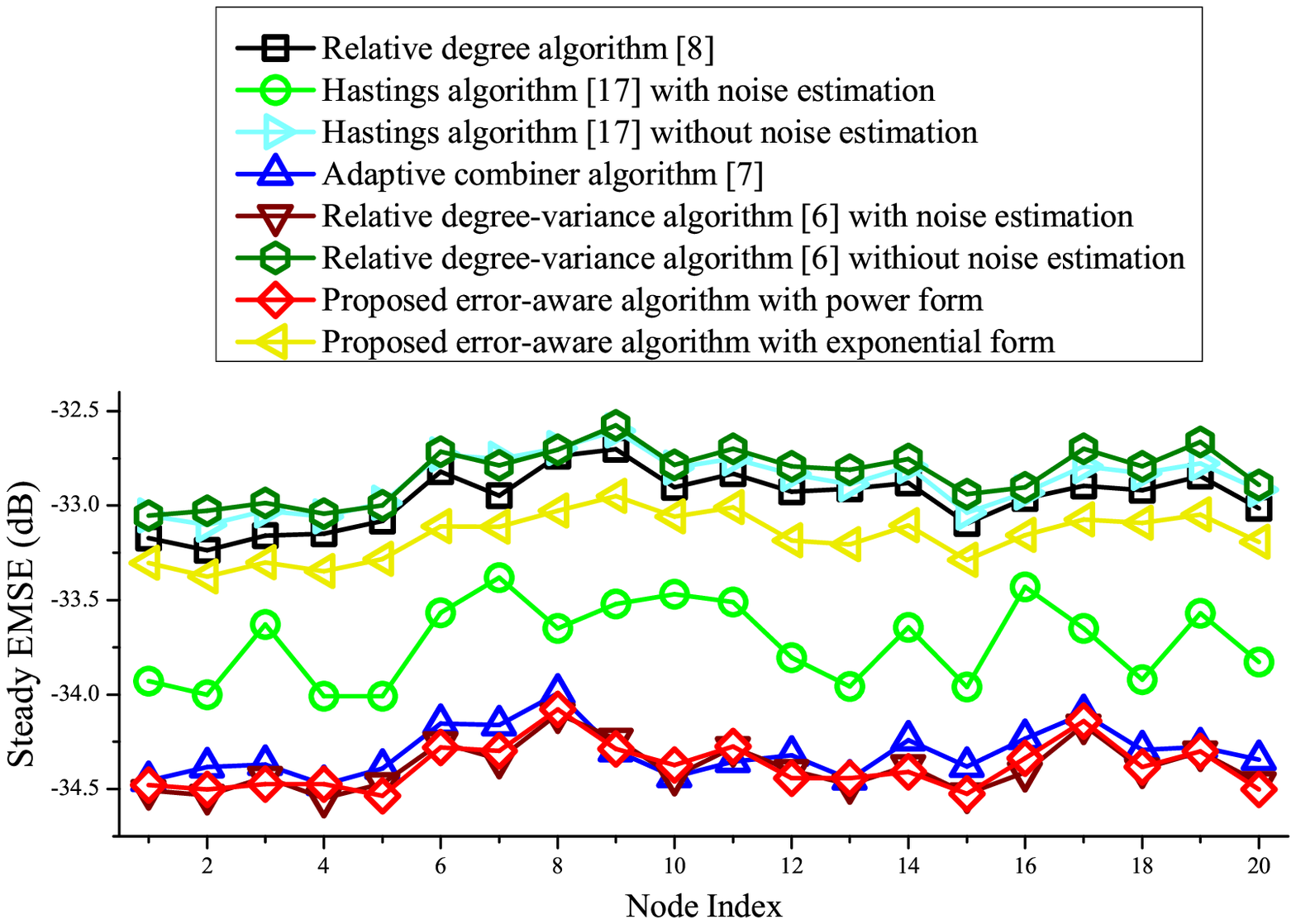,width=7.5cm}}
  \centerline{\scriptsize{(a) Node's EMSE.}}\vspace{0.1cm}
\end{minipage}
\begin{minipage}[b]{1\linewidth}
  \centering
  \centerline{\epsfig{figure=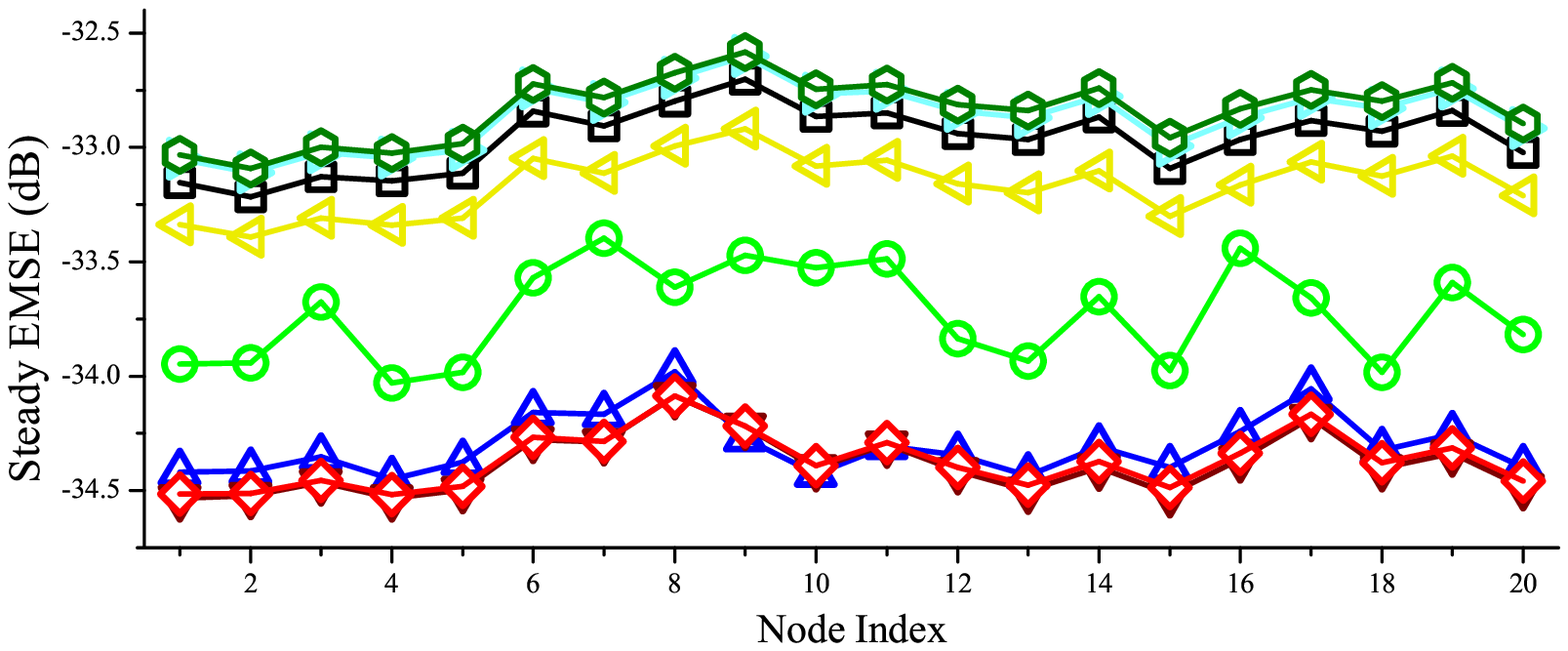,width=7.5cm}}
  \centerline{\scriptsize{(b) Node's MSD.}}
\end{minipage}
\caption{Steady performances comparison with unknown noise variances.}\label{vary}
\end{figure}

In the second comparison, we assume that the noise variance of each node is unknown, but can be estimated by the method proposed in \cite{two}. Fig.\,\ref{netmsev} and Fig.\,\ref{vary} show the transient and steady-state performances for six kinds of algorithms in terms of EMSE and MSD under similar convergence rate. Since the noise variance estimation requires additional complexity, we also simulate the Hastings and relative degree-variance algorithms without variance estimation for fair comparison, where the noise variance is set as the network average variance, which is assumed to be prior information. Comparing with Fig.\,\ref{steadynetmse}, we can see that when the noise variance information is not available, the performance degradation of relative degree-variance algorithm is significant, about 0.5dB (12\% more error) even with noise variance estimation, while the performance of Hastings algorithm degrades only a little since it relies less on the noise variance information. From Fig.\,\ref{netmsev}-(b), we can clearly see that when the variance estimation method is not adopted, our proposed algorithm with power form achieves the best performance. When the variance estimation method is adopted, the performances of our proposed algorithm with power form, the relative degree-variance and the adaptive combiner algorithm are similar, all of which perform better than other algorithms. Nevertheless, the complexity of both relative degree-variance algorithm with variance estimation and the adaptive combiner algorithm are higher than that of our proposed algorithm with power form. Such results immediately show the advantage of the proposed general framework. We should notice that more algorithms with better performances under certain criteria can be designed based on the proposed framework by choosing more proper fitness functions.


\subsection{Diffusion Probability}

\begin{figure}[!t]
  \centering
\begin{minipage}[b]{1\linewidth}
  \centering
  \centerline{\epsfig{figure=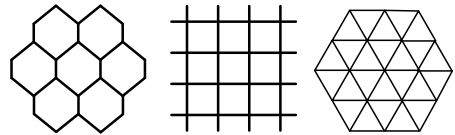,width=6cm}}
  \centerline{\scriptsize{(a) Regular graph structures with degree $n=3$, $4$ and $6$.}}\vspace{0.5cm}
\end{minipage}
\begin{minipage}[b]{1\linewidth}
  \centering
  \centerline{\epsfig{figure=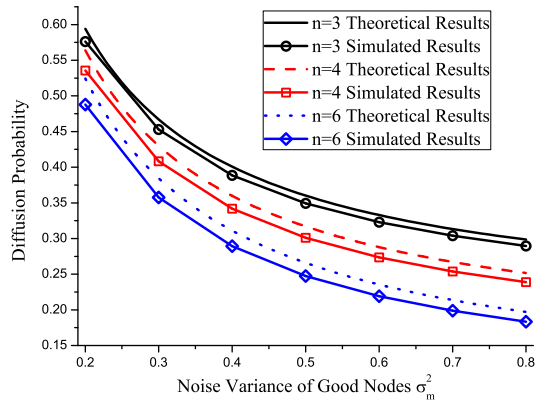,width=7cm}}
  \centerline{\scriptsize{(b) Diffusion probability.}}
\end{minipage}
\caption{Diffusion probabilities under three types of regular graphs.}\label{diffusionfig}
\end{figure}

In this subsection, we develop simulation to verify the diffusion probability analysis in Section IV. For the simulation setup, three types of regular graphs are generated with degree $n=3$, $4$ and $6$, respectively, as shown in Fig.\,\ref{diffusionfig}-(a). All these three types of graphs are with $N=100$ nodes, where each node's trace of regressor covariance is set to be $\mbox{Tr}(\bm R_u)=10$, the common nodes's noise variance is set as $\sigma^2_r=1.5$ and the good node's noise variance is set as $\sigma^2_m\in[0.2,0.8]$. In the simulation, the network is initialized with the state that all common nodes choosing strategy $\bm S_r$. Then, at each time step, a randomly chosen node's strategy is updated according to the IM rules under weak selection ($w = 0.01$), as illustrated in Section III-B. The update steps are repeated until either strategy $\bm S_m$ has reached fixation or the number of steps has reach the limit. The diffusion probability is calculated by the fraction of runs where strategy $\bm S_m$ reached fixation out of $10^6$ runs. Fig.\,\ref{diffusionfig}-(b) shows the simulation results, from which we can see that all the simulated results are basically accord with the corresponding theoretical results and the gaps are due to the approximation during the derivations. Moreover, we can see that the diffusion probability of good signal decreases along with the increase of its noise variance, i.e., better signal has better diffusion capability.

\subsection{Evolutionarily Stable Strategy}

\begin{figure*}[!t]
    \centerline{\epsfig{figure=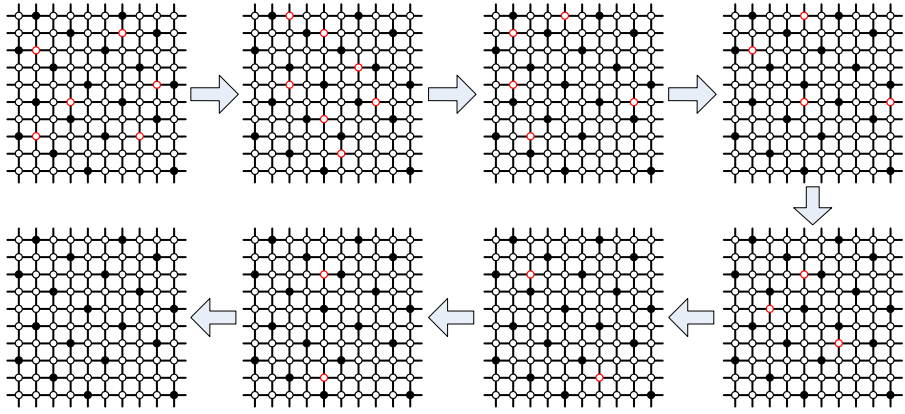,width=12cm}}
    \caption{Strategy updating process in a $10\times10$ grid network with degree $n=4$ and number of nodes $N=100$.}\label{essfig}
\end{figure*}

To verify that strategy $\bm S_m$ is an ESS in the adaptive network, we further simulate the IM update rule on a $10\times10$ grid network with degree $n=4$ and number of nodes $N=100$, as shown in Fig.\,\ref{essfig} where the hollow points represent common nodes and the solid nodes represent good nodes. In the simulation, all the settings are same with those in the simulation of diffusion probability in Section VI-B, except the initial network setting. The initial network state is set that the majority of nodes adopt strategy $\bm S_m$ denoted with black color (including both hollow and solid nodes) in Fig.\,\ref{essfig}, and only a very small percentage of nodes use strategy $\bm S_r$ denoted with red color. From the strategy updating process of the whole network illustrated in Fig.\,\ref{essfig}, we can see that the network finally abandons the unfavorable strategy $\bm S_r$, which verifies the stability of strategy $\bm S_m$. 

\section{Conclusion}

In this paper, we proposed an evolutionary game theoretic framework to offer a very general view of the distributed adaptive filtering problems and unify existing algorithms. Based on this framework, as examples, we further designed two error-aware adaptive filtering algorithms. Using the graphical evolutionary game theory, we analyzed the information diffusion process in the network under the IM update rule, and proved that the strategy of using information from nodes with good signal is always an ESS. We would like to emphasize that, unlike the traditional bottom-up approaches, the proposed graphical evolutionary game theoretic framework provides a top-down design philosophy to understand the fundamentals of distributed adaptive algorithms. Such a top-down design philosophy is very important to the field of distributed adaptive signal process, since it offers a unified view of the formulation and can inspire more new distributed adaptive algorithms to be designed in the future.

\appendix[Proof of Theorem 1]
\begin{proof}
First, let us define $m(p_m)$ as the mean of the increment of $p_m$ per unit time given as follows
\begin{eqnarray}
\!\!\!\!\!\!m(p_m)\!\!\!\!&=&\!\!\!\! \frac{\dot p_m}{1/N}\nonumber\\
\!\!\!\!&\simeq&\!\!\!\!\frac{\alpha n(n-2)}{(n-1)(n+1)^2}p_m(1-p_m)(ap_m+b).
\end{eqnarray}
where the second step is derived by substituting (\ref{steady})-(\ref{steady2}) into (\ref{dpm}) and the parameters $a$ and $b$ are given as follows:
\begin{eqnarray}
\!\!\!\!\!\!\!\!a\!\!\!\!\!&=&\!\!\!\!(n-2)(n+3)(u_1-u_2-u_3+u_4),\label{a}\\
\!\!\!\!\!\!\!\!b\!\!\!\!\!&=&\!\!\!\!-(n\!-\!1)(n\!+\!3)u_1\!-\!3u_2\!+\!(n^2\!+\!n\!-\!3)u_3\!+\!(n\!+\!3)u_4.\label{b}
\end{eqnarray}
We then define $v(p_m)$ as the variance of the increment of $p_m$ per unit time, which can be calculated by
\begin{equation}
v(p_m)= \frac{\dot {p^2_m}- (\dot p_m)^2}{1/N},\label{vpm}\\
\end{equation}
where $\dot {p^2_m}$ can be computed by
\begin{eqnarray}
\dot {p^2_m}\!\!\!\!&=&\!\!\!\!\frac{1}{N^2}\Bigg(\mbox{Prob}\Big(\Delta p_m=\frac{1}{N}\Big)+\mbox{Prob}\Big(\Delta p_m=-\frac{1}{N}\Big)\Bigg)\nonumber\\
\!\!\!\!&=&\!\!\!\!\frac{2}{N^2}\frac{n(n-2)}{(n-1)(n+1)}p_m(1-p_m)+O(\alpha).\label{p2m}
\end{eqnarray}
In such a case, $v(p_m)$ can be approximated by
\begin{equation}
v(p_m)\simeq\frac{2}{N}\frac{n(n-2)}{(n-1)(n+1)}p_m(1-p_m).\\
\end{equation}
Suppose the initial percentage of good nodes in the network is $p_{m0}$. Let us define $H(p_{m0})$ as the probability that these good signals can finally be adopted by the whole network, i.e., all nodes can update their own $\bm w$ using information from good nodes. According to the backward Kolmogorov equation \cite{kol}, $H(p_{m0})$ satisfies following differential equation
\begin{equation}
0=m(p_{m0})\frac{dH(p_{m0})}{dp_{m0}}+\frac{v(p_{m0})}{2}\frac{d^2H(p_{m0})}{dp^2_{m0}}.
\end{equation}
With the weak selection assumption, we can have the approximate solution of $H(p_{m0})$ as
\begin{equation}
H(p_{m0})=p_{m0}+\frac{\alpha N}{6(n+1)}p_{m0}(1-p_{m0})\Big((a+3b)+ap_{m0}\Big).\label{fixation}
\end{equation}

Let us consider the worst initial system state that each common node has connection with only one good node, i,e., $p_{m0}=\frac{1}{n+1}$, we have
\begin{equation}
H\bigg(\frac{1}{n+1}\bigg)\simeq\frac{1}{n+1}+\frac{\alpha nN}{6(n+1)^3}(a+3b).\label{probability}
\end{equation}
By substituting (\ref{a}) and (\ref{b}) into (\ref{probability}), we can have the close-form expression for the diffusion probability in (\ref{difc}). This completes the proof of the theorem.
\end{proof}

\textbf{Remark:} From (\ref{fixation}), we can see that there are two terms constituting the expression of diffusion probability: the initial percentage of strategy $\bm S_m$, $p_{m_0}$ (the initial system state) and the second term representing the changes of system state after beginning, in which $a+3b$ determines whether $p_m$ is increasing or decreasing along with the system updating. If $a+3b<0$, i.e., the diffusion probability is even lower than the initial percentage of strategy $\bm S_m$, the information from good nodes are shrinking over the network, instead of spreading. Therefore, $a+3b>0$ is more favorable for the improvement of the adaptive network performance.
\bibliographystyle{IEEEtran}

\bibliography{list}

\begin{IEEEbiography}[{\includegraphics[width=1in,height=1.25in,clip,keepaspectratio]{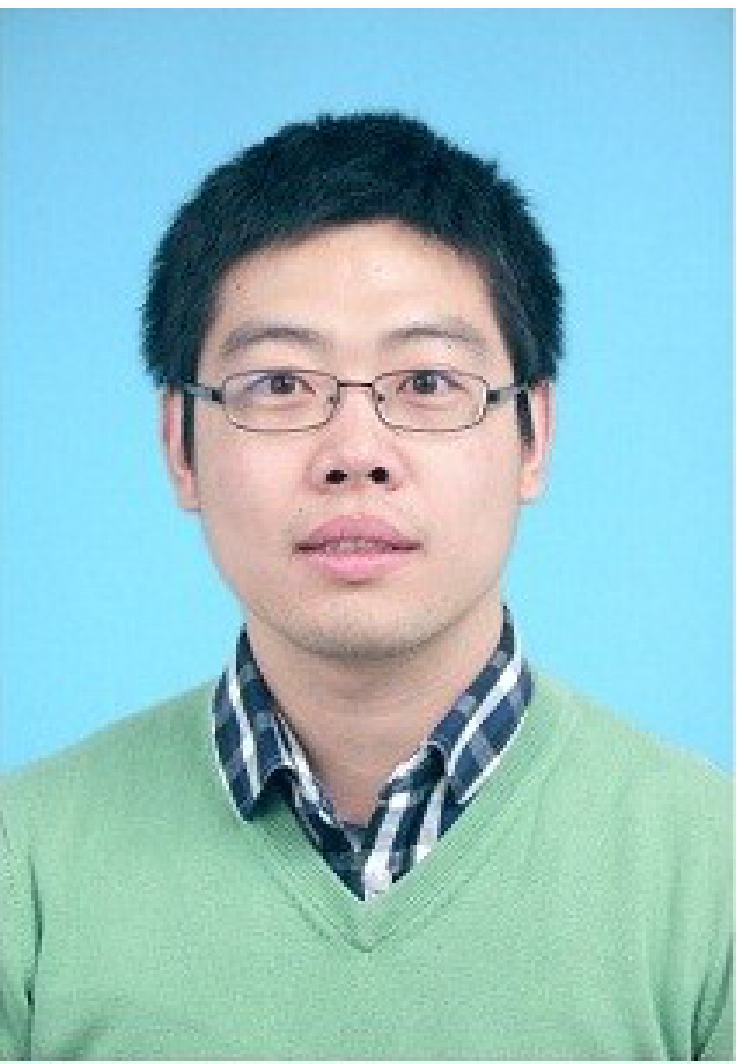}}]{Chunxiao Jiang}(S'09-M'13)
received his B.S. degree in information engineering from Beijing University of Aeronautics and Astronautics (Beihang University) in 2008 and the Ph.D. degree from Tsinghua
University (THU), Beijing in 2013, both with
the highest honors. During 2011-2012, he visited the
Signals and Information Group (SIG) at Department
of Electrical \& Computer Engineering (ECE) of
University of Maryland (UMD), supported by China
Scholarship Council (CSC) for one year. 

Dr. Jiang is
currently a research associate in ECE department of
UMD with Prof. K. J. Ray Liu, and also a post-doctor in EE department
of THU. His research interests include the applications of game theory
and queuing theory in wireless communication and networking and social
networks.

Dr. Jiang received the Beijing Distinguished Graduated Student Award, Chinese
National Fellowship and Tsinghua Outstanding Distinguished Doctoral
Dissertation in 2013.
\end{IEEEbiography}

\begin{IEEEbiography}[{\includegraphics[width=1in,height=1.25in,clip,keepaspectratio]{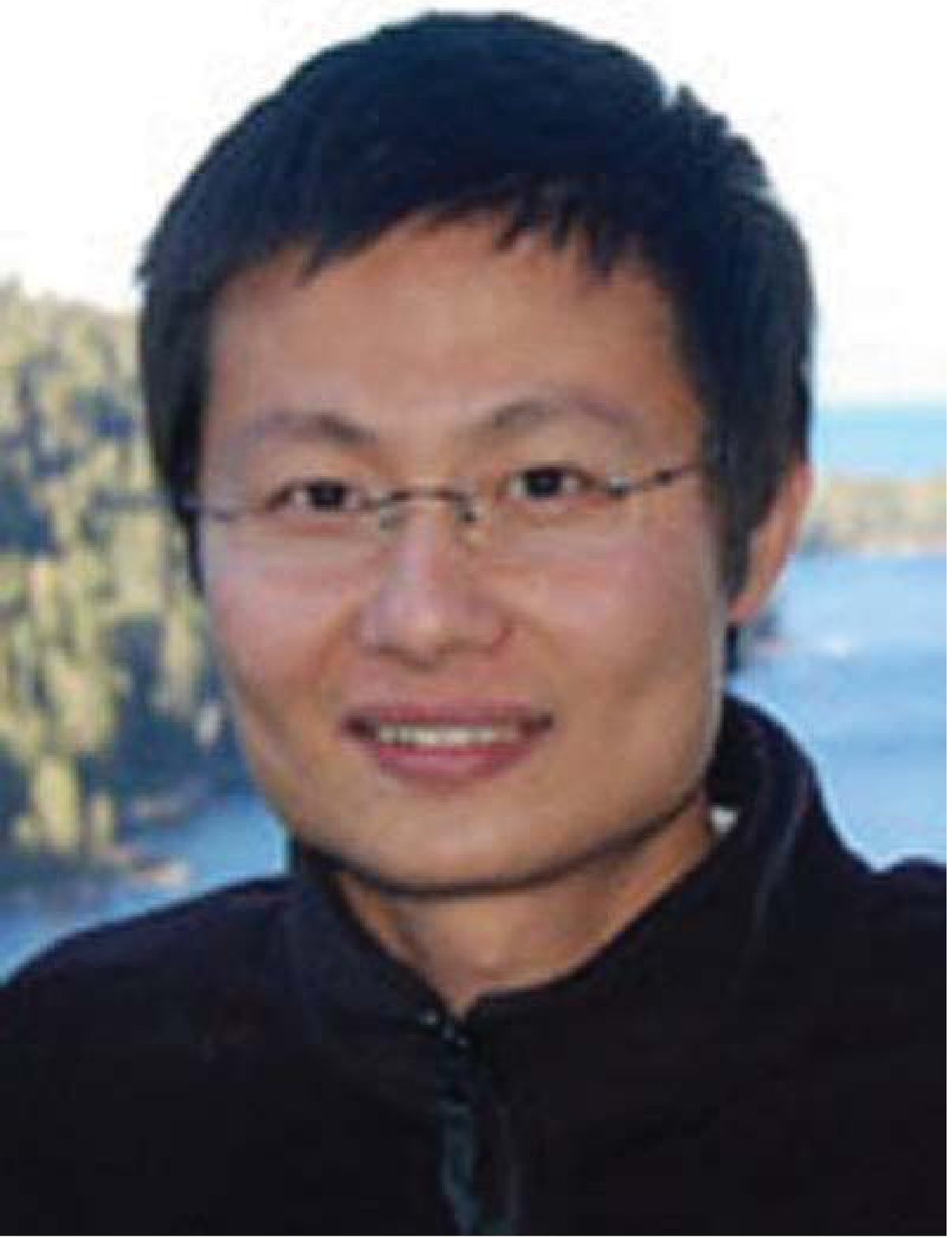}}]{Yan Chen}(S'06-M'11)
received the Bachelor's degree from University of Science and Technology of China in 2004, the M. Phil degree from Hong Kong University of Science and Technology (HKUST) in 2007, and the Ph.D. degree from University of Maryland College Park in 2011. From 2011 to 2013, he is a Postdoctoral research associate in the Department of Electrical and Computer Engineering at University of Maryland College Park.

Currently, he is a Principal Technologist at Origin Wireless Communications. He is also affiliated with Signal and Information Group of University of Maryland College Park. His current research interests are in social learning and networking, behavior analysis and mechanism design for network systems, multimedia signal processing and communication.

Dr. Chen received the University of Maryland Future Faculty Fellowship
in 2010, Chinese Government Award for outstanding students abroad in
2011, University of Maryland ECE Distinguished Dissertation Fellowship
Honorable Mention in 2011, and was the Finalist of A. James Clark School
of Engineering Deans Doctoral Research Award in 2011.
\end{IEEEbiography}

\begin{IEEEbiography}[{\includegraphics[width=1in,height=1.25in,clip,keepaspectratio]{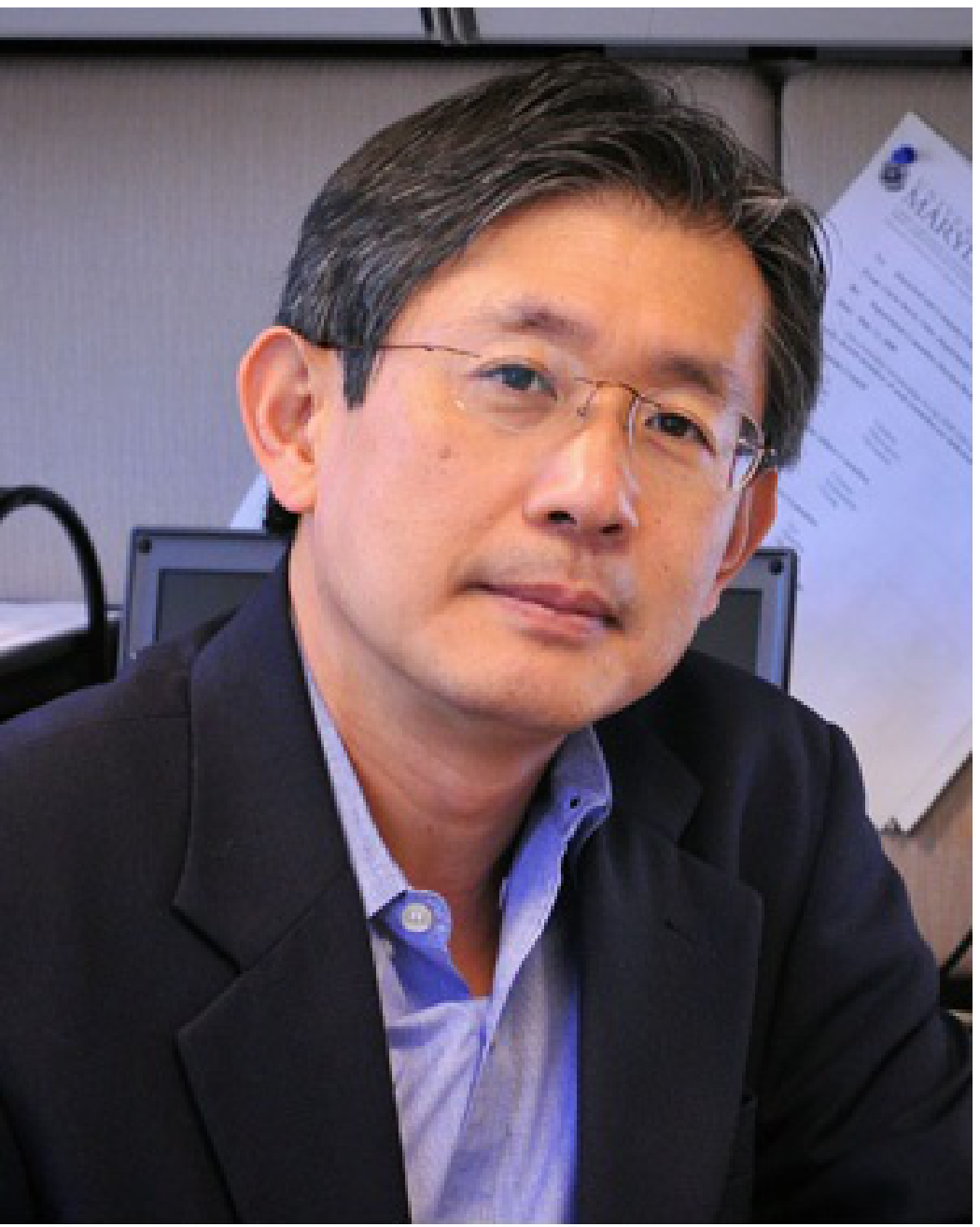}}]{K. J. Ray Liu}(F'03)
was named a Distinguished Scholar-Teacher of University of Maryland, College Park, in 2007, where he is Christine Kim Eminent Professor of Information Technology. He leads the
Maryland Signals and Information Group conducting research encompassing broad areas of signal processing and communications with recent focus on cooperative and cognitive communications, social learning and network science, information forensics and security, and green information and communications technology.

Dr. Liu is the recipient of numerous honors and awards including IEEE Signal Processing Society Technical Achievement Award and Distinguished Lecturer. He also received various teaching and research recognitions from University of Maryland including university-level Invention of the Year Award; and Poole and Kent Senior Faculty Teaching Award, Outstanding Faculty Research Award, and Outstanding Faculty Service Award, all from A. James Clark School of Engineering. An ISI Highly Cited Author, Dr. Liu is a Fellow of IEEE and AAAS.

Dr. Liu is President of IEEE Signal Processing Society where he has served as Vice President -- Publications and Board of Governor. He was the Editor-in-Chief of IEEE Signal Processing Magazine and the founding Editor-in-Chief of EURASIP Journal on Advances in Signal Processing.
\end{IEEEbiography}

\end{document}